\pdfoutput=1  
\documentclass[]{article}  
\usepackage{url,float}
\usepackage{graphicx}
\usepackage{amsfonts}
\usepackage{amssymb}
\usepackage{latexsym}

\newcommand{\hide}[1]{}

\newcommand{\ABox}{
\raisebox{3pt}{\framebox[6pt]{\rule{6pt}{0pt}}}
}
\newenvironment{proof}{{\bf Proof:}}{\hfill\ABox}

\newtheorem{theorem}{{\bf Theorem}}

\newtheorem{lemma}{Lemma}

\newcommand{\lemlab}[1]{\label{lemma:#1}}
\newcommand{\thmlab}[1]{\label{thm:#1}}

\newcommand{\figlab}[1]{\label{fig:#1}}
\newcommand{\seclab}[1]{\label{sec:#1}}

\newcommand{\lemref}[1]{\ref{lemma:#1}}
\newcommand{\thmref}[1]{\ref{thm:#1}}

\newcommand{\secref}[1]{\ref{sec:#1}}
\newcommand{\eqref}[1]{\ref{eq:#1}}
\newcommand{\figref}[1]{\ref{fig:#1}}

{\makeatletter
 \gdef\xxxmark{%
   \expandafter\ifx\csname @mpargs\endcsname\relax 
     \expandafter\ifx\csname @captype\endcsname\relax 
       \marginpar{xxx}
     \else
       xxx 
     \fi
   \else
     xxx 
   \fi}
 \gdef\xxx{\@ifnextchar[\xxx@lab\xxx@nolab}
 \long\gdef\xxx@lab[#1]#2{{\bf [\xxxmark #2 ---{\sc #1}]}}
 \long\gdef\xxx@nolab#1{{\bf [\xxxmark #1]}}
 \gdef\turnoffxxx{\long\gdef\xxx@lab[##1]##2{}\long\gdef\xxx@nolab##1{}}%
}

\def\P{{\mathcal P}}

\def\X{{\mathcal X}}
\def\G{{\Gamma}}
\def\g{{\gamma}}

\def\o{{\omega}}

\def\s{{\sigma}}

\def\a{{\alpha}}
\def\b{{\beta}}
\def\t{{\theta}}
\def\sp{\mathop{\rm sp}\nolimits}
\def\bP{{\partial P}}
\def\bX{{\partial X}}

\def\bU{{\partial U}}
\def\bM{{\partial M}}

\newcommand{\squeezelist}{\setlength{\itemsep}{0pt}}

\title{
Star Unfolding Convex Polyhedra \\ 
via
Quasigeodesic Loops\footnote{
   A preliminary version of this work appeared
   in~\cite{iov-ucpq-07,iov-ucpq-07a}.
}
}

\author{%
Jin-ichi Itoh%
    \thanks{Department of Mathematics,
        Faculty Education, Kumamoto University,
        Kumamoto 860-8555, Japan.
    \protect\url{j-itoh@kumamoto-u.ac.jp}}
\and
Joseph O'Rourke%
    \thanks{Department of Computer Science, Smith College, Northampton, MA
      01063, USA.
      \protect\url{orourke@cs.smith.edu}.}
\and
Costin V\^{i}lcu%
    \thanks{Institute of Mathematics ``Simion Stoilow'' Romanian Academy,
      P.O. Box 1-764,
      RO-014700 Bucharest, Romania.
    \protect\url{Costin.Vilcu@imar.ro}.
    Partially supported by Romanian Government grant PN II Idei 1187.
}
}

\begin{document}
\maketitle

\begin{abstract}
We extend the notion of star unfolding to be based on
a 
quasigeodesic loop $Q$ rather than on a point.
This gives a new general method to unfold the surface of any convex polyhedron
$\P$ to a simple (non-overlapping), planar polygon: 
cut along one shortest path from each vertex of $\P$ to $Q$, 
and cut all but one segment of $Q$.
\end{abstract}

\section{Introduction}
\seclab{Introduction}

There are two general methods known to unfold the surface $\P$ 
of any convex polyhedron
to a simple (non-overlapping) polygon in the plane:
the source unfolding and the star unfolding.
Both unfoldings are with respect to a point $x \in \P$.
Here we define a third general method:
the star unfolding
with respect to a
simple closed ``quasigeodesic loop'' $Q$ on $\P$.
In a companion paper~\cite{iov-sucpr-09}, we 
extend the analysis
to the source unfolding with respect to a wider class of curves $Q$.

The \emph{point source unfolding} cuts the \emph{cut locus} of
the point $x$:
the closure of the set of all those points $y$ to which there is
more than one shortest path on $\P$ from $x$.
The notion of cut locus was introduced by 
Poincar\'e~\cite{p-lgsc-1905}
in 1905, and since then has gained an
important place in global Riemannian geometry; see, e.g., 
\cite{k-ccl-67} or~\cite{s-rg-96}.
The point source unfolding has been studied
for polyhedral convex surfaces since~\cite{ss-spps-86} 
(where the cut locus is called the ``ridge tree'').

The \emph{point star unfolding} cuts the shortest paths from $x$ to every
vertex of $\P$.
The idea goes back to Alexandrov~\cite[p.~181]{a-kp-48};%
\footnote{
   It is called the ``Alexandrov unfolding'' in~\cite{mp-mccpc-05}.
}
that it unfolds $\P$ to a simple (non-overlapping) polygon was established
in~\cite{ao-nsu-92}.

In this paper we extend the star unfolding to be based on a
simple closed polygonal curve $Q$ with particular properties,
rather than 
on a single point.
This unfolds any convex polyhedron to a simple polygon,
answering a question
raised in~\cite[p.~307]{do-gfalop-07}.

The curves $Q$ for which our star unfolding works are
quasigeodesic loops, which we now define.

\paragraph{Geodesics \& Quasigeodesics.}
Let $\G$ be any directed curve on a convex surface $\P$, 
and $p \in \G$ be any point in the relative interior of $\G$, i.e., not
an endpoint.
Let $L(p)$ be the total surface angle incident to the left side of $p$,
and $R(p)$ the angle to the right side.
$\G$ is a \emph{geodesic} if $L(p){=}R(p)=\pi$.
A \emph{quasigeodesic} $\G$ loosens this condition to $L(p) \le \pi$ and $R(p) \le \pi$,
again for all $p$ interior to $\G$~\cite[p.~16]{az-igs-67}~\cite[p.~28]{p-egcs-73}.
So a quasigeodesic $\G$ has $\pi$ total face angle incident to each
side at all nonvertex points (just like a geodesic),
and has at most $\pi$ angle to each side where $\G$ passes through a
polyhedron vertex.
A \emph{simple closed geodesic} is non-self-intersecting (\emph{simple}) closed
curve that is a geodesic, and
a \emph{simple closed quasigeodesic} is a simple closed curve
on $\P$ that is quasigeodesic throughout its length.
As all curves we consider must be simple, we will henceforth
drop that prefix.

A \emph{geodesic loop} is a closed curve that is geodesic
everywhere except possibly at one point,
and similarly a \emph{quasigeodesic loop} is quasigeodesic except
possibly at one point $x$, the \emph{loop point}, at which the angle conditions
on $L(x)$ and $R(x)$ may be violated---one may be larger than $\pi$.
Quasigeodesic loops encompass
closed geodesics and quasigeodesics,
as well as geodesic loops.

Pogorelov showed that any convex polyhedron $\P$ has at least three closed
quasigeodesics~\cite{p-qglcs-49}, extending the celebrated earlier result of
Lyusternik-Schnirelmann showing the same for
differentiable convex surfaces.
However,
there is no algorithm known that will find a simple closed quasigeodesic
in polynomial time:
Open Problem~24.2~\cite[p.~374]{do-gfalop-07}.

Fortunately it is in general easy to find quasigeodesic loops on a given $\P$:
start at any nonvertex point $p$, and extend a geodesic from
$p$ in opposite directions, following each branch until a 
self-intersection point is found, either between branches or within one branch.
If no vertices are encountered, we have a geodesic loop;
if vertices are encountered, at each vertex continue in an arbitrary direction
that maintains quasigeodesicity, 
to obtain a quasigeodesic loop. 
An exception to this ease of finding a quasigeodesic loop could
occur on an \emph{isosceles tetrahedron}: a tetrahedron
whose four faces are congruent triangles, or, equivalently,
one at which the total face angle incident to each vertex is $\pi$.
It is proved in~\cite{iv-gcit-08} that a convex surface possesses
a \emph{simple quasigeodesic line}---a
non-self-intersecting quasigeodesic infinite in both directions---if
and only if the surface is an isosceles tetrahedron.

\paragraph{Discrete Curvature.}
The discrete \emph{curvature} $\o(p)$ at any point $p \in \P$
is the \emph{angle deficit}
or \emph{gap}: $2\pi$ minus the sum of the face angles incident to $p$.
The curvature is only nonzero at vertices of $\P$;
at each vertex it is positive because $\P$ is convex.
By the Gauss-Bonnet theorem, a closed geodesic partitions the
curvature into $2\pi$ in each ``hemisphere'' of $P$.
For quasigeodesics that pass through vertices, the curvature in each
half is at most $2\pi$.
The curvature in each half defined by a quasigeodesic loop depends on the
angle at the loop point.

\paragraph{Some Notation.}
For a quasigeodesic loop $Q$ on $\P$,
$\P \setminus Q$ separates $\P$ into two ``halves''
$P_1$ and $P_2$.
As our main focus is usually on one such half, 
to ease notation we sometimes use $P$ without a subscript
to represent either of $P_1$ or $P_2$
when the distinction between them does not matter.
Unless otherwise stated,
vertices of $\P$ are labeled $v_i$ in arbitrary order.
We will use $p p'$ to denote a shortest path on $\P$ between $p$ and $p'$.
Other notation will be introduced as needed.


\section{Examples and Algorithm}
\seclab{Example}
We start with an example.
Figure~\figref{GeodesicLoopCube}(a) shows a geodesic loop
$Q$ on the surface $\P$ of a cube.
$L(p){=}R(p)=\pi$ at every point $p$ of $Q$ except at
$x$, where $R(x){=}\frac{3}{2}\pi$ and  $L(x){=}\frac{1}{2}\pi$.
Note that three cube vertices, 
$v_3, \; v_6, \; v_7$
are to the left
of $Q$, and the other five to the right.
This is consistent with the Gauss-Bonnet theorem, because
$Q$ has a total turn of $\frac{1}{2}\pi$,
so turn plus enclosed curvature is $2\pi$.

For each vertex $v_i \in \P$, we select a shortest path
$\sp(v_i)=v_i v'_i$ to $Q$: 
a geodesic from $v_i$ to a point $v'_i \in Q$
whose length is minimal among all geodesics to $Q$.
In general there could be several shortest paths from $v_i$ to $Q$;
we use $\sp(v_i)$ to represent an arbitrarily selected one.
The point $v'_i \in Q$ is called
a \emph{projection} of $v_i$ onto $Q$.
In this example, for each $v_i$ there is a unique shortest path $\sp(v_i)$,
which is the generic situation.

\begin{figure}[htbp]
\centering
\includegraphics[width=\linewidth]{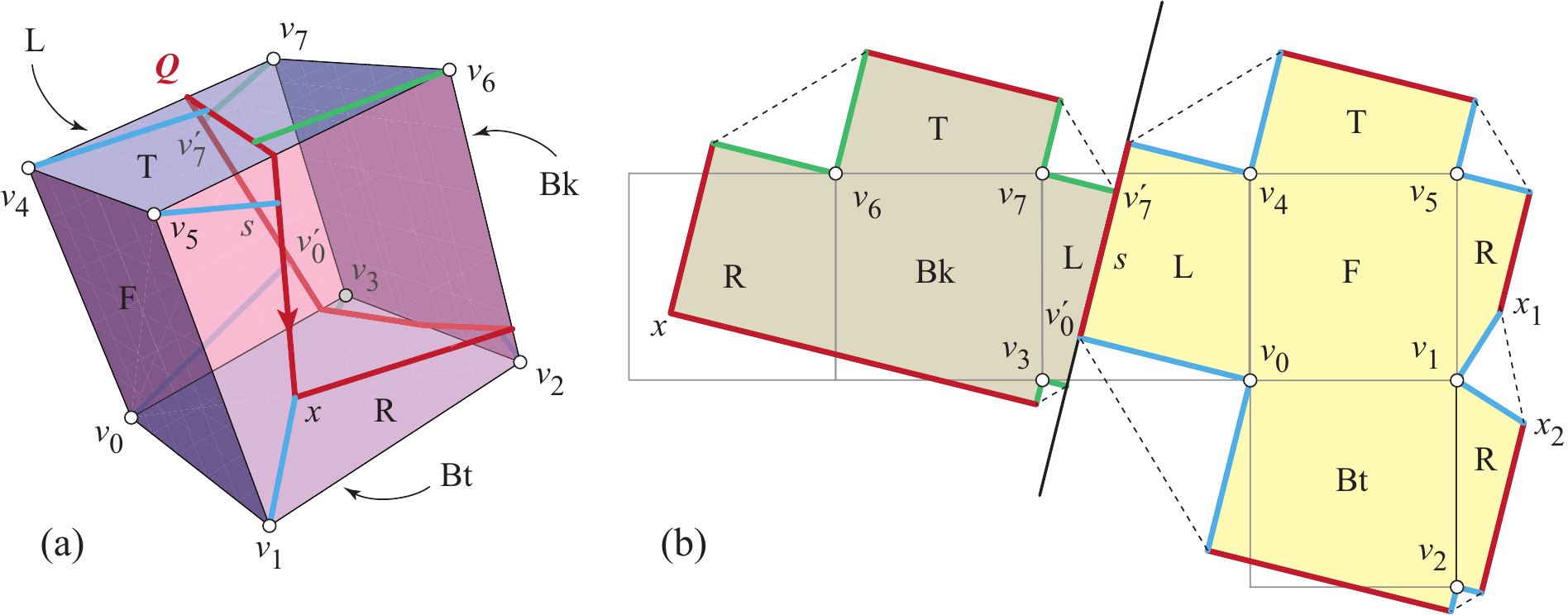}
\caption{
(a)~Geodesic loop $Q$ on cube.
Shortest paths $\sp(v_i)$ are shown.  Faces are labeled
$\{F,T,L,R,Bt,Bk\}$.
(b)~Star unfolding with respect to $Q$,
joined at 
$s=v'_0 v'_7$.
}
\figlab{GeodesicLoopCube}
\end{figure}

\paragraph{Algorithm.}
If we view the star unfolding as an algorithm
with inputs $\P$ and $Q$, it consists of three
main steps:
\begin{enumerate}
\squeezelist
\item Select shortest paths $\sp(v_i)$ from each $v_i \in \P$ to $Q$.
\item Cut along all $\sp(v_i)$ and flatten each half.
\item Cut along $Q$, joining the two halves at an uncut segment $s \subset Q$.
\end{enumerate}

After cutting along $\sp(v_i)$, we conceptually insert an isosceles
triangle with apex angle $\o(v_i)$ at each $v_i$, which flattens
each resulting half.  One half (in our example, the left half), is convex,
while the other resulting half may have several points of nonconvexity,
at the images 
of $x$.
(In our example, only the image $x_1$ is nonconvex, when the inserted
``curvature triangles'' are included.)
In the third and final step of the procedure,
we select a segment $s$ of $Q$ whose interior contains
neither a vertex $v_i$ nor any vertex projection $v'_i$, 
such that the extension of $s$ is a supporting line of each half,
and cut all of $Q$ except for $s$.
In our first example, we choose 
$s=v'_0 v'_7$
(many choices for $s$ work in this example),
which leads to non-overlap of the two halves.

We illustrate with one more example before proceeding.
Again $\P$ is a cube, but now $Q$ is the closed quasigeodesic composed
of the edges bounding one square face,
the bottom face in
Figure~\figref{LatinCross}(a).
Cutting four shortest paths from the other vertices orthogonal
to $Q$, and cutting all but one edge $s$ of $Q$, results
in the standard Latin-cross unfolding of the cube shown in~(b).
\begin{figure}[htbp]
\centering
\includegraphics[width=0.75\linewidth]{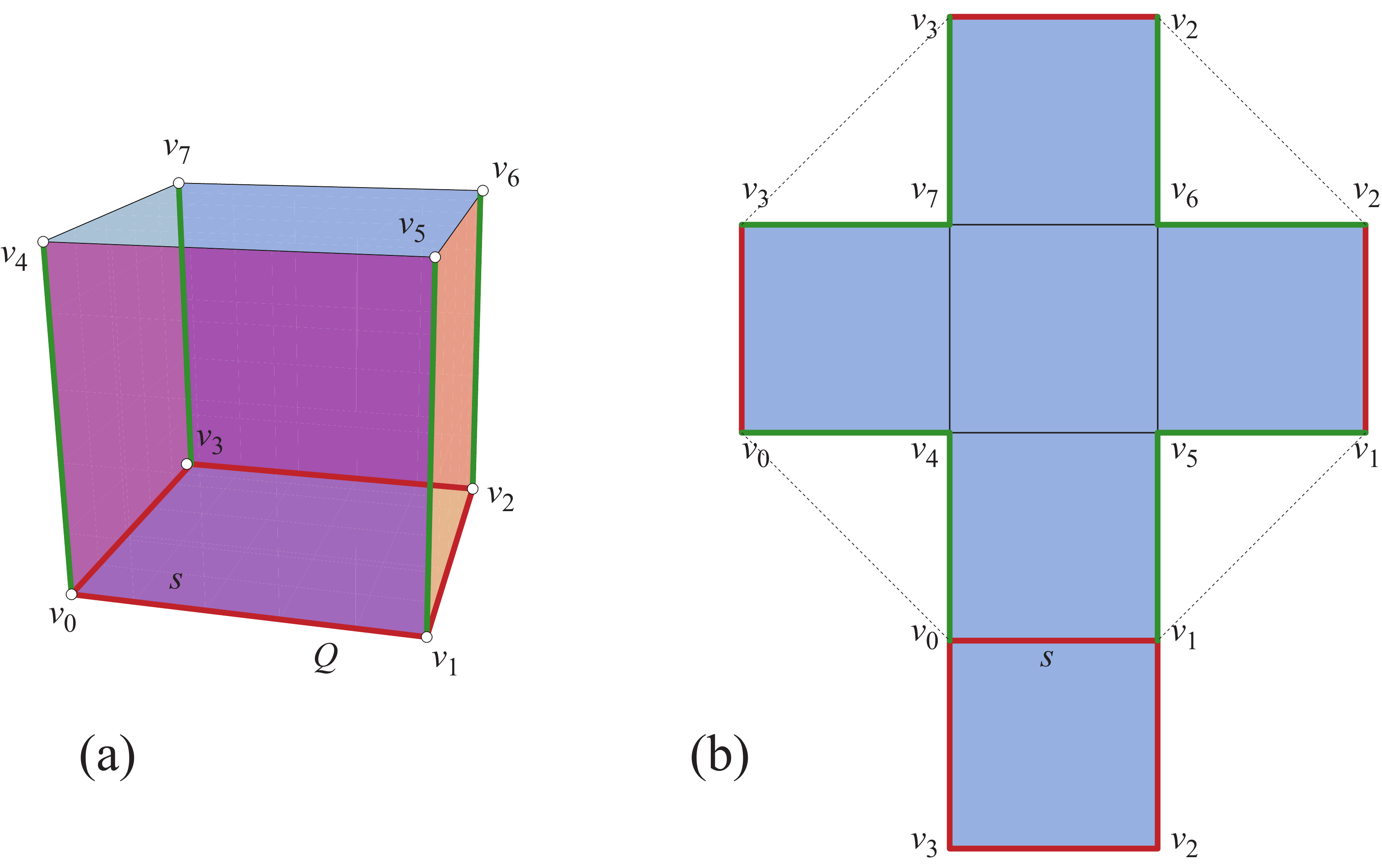}
\caption{(a)~$Q=(v_0,v_1,v_2,v_3)$.
(b)~Star unfolding with respect to $Q$.
Here the uncut segment of $Q$ is $s= v_0 v_1$.
}
\figlab{LatinCross}
\end{figure}

We now proceed to detail the three steps of the procedure,
this time with proofs.  
Because the proofs for quasigeodesics are straightforward in comparison
to the proofs for quasigeodesic loops, we separate the two
in the exposition.

\section{Quasigeodesics}

\subsection{Shortest Path Cuts for  Quasigeodesics}
We again use
a cube as an illustrative example, but this time with a 
closed quasigeodesic $Q$, not a loop: 
$Q=(v_0,v_5,v_7)$; see
Figure~\figref{cube_geodesic_hemis}(a).
There is $\pi$ angle incident to the right at $v_5$, 
and $\frac{1}{2}\pi$ incident to
the left; and similarly at $v_0$ and $v_7$.  At all other
points $p \in Q$, $L(p){=}R(p)=\pi$.  Thus $Q$ is indeed a quasigeodesic.
We will call the right half (including $v_2$) $P_1$, and the left half 
(including $v_4$) $P_2$.
In Figure~\figref{cube_geodesic_hemis}(a), the paths from $\{v_1,v_3,v_6\}$
are uniquely shortest.
From $v_2$ there are three paths tied for shortest, and from
$v_4$ also three are tied.
\begin{figure}[htbp]
\centering
\includegraphics[width=0.75\linewidth]{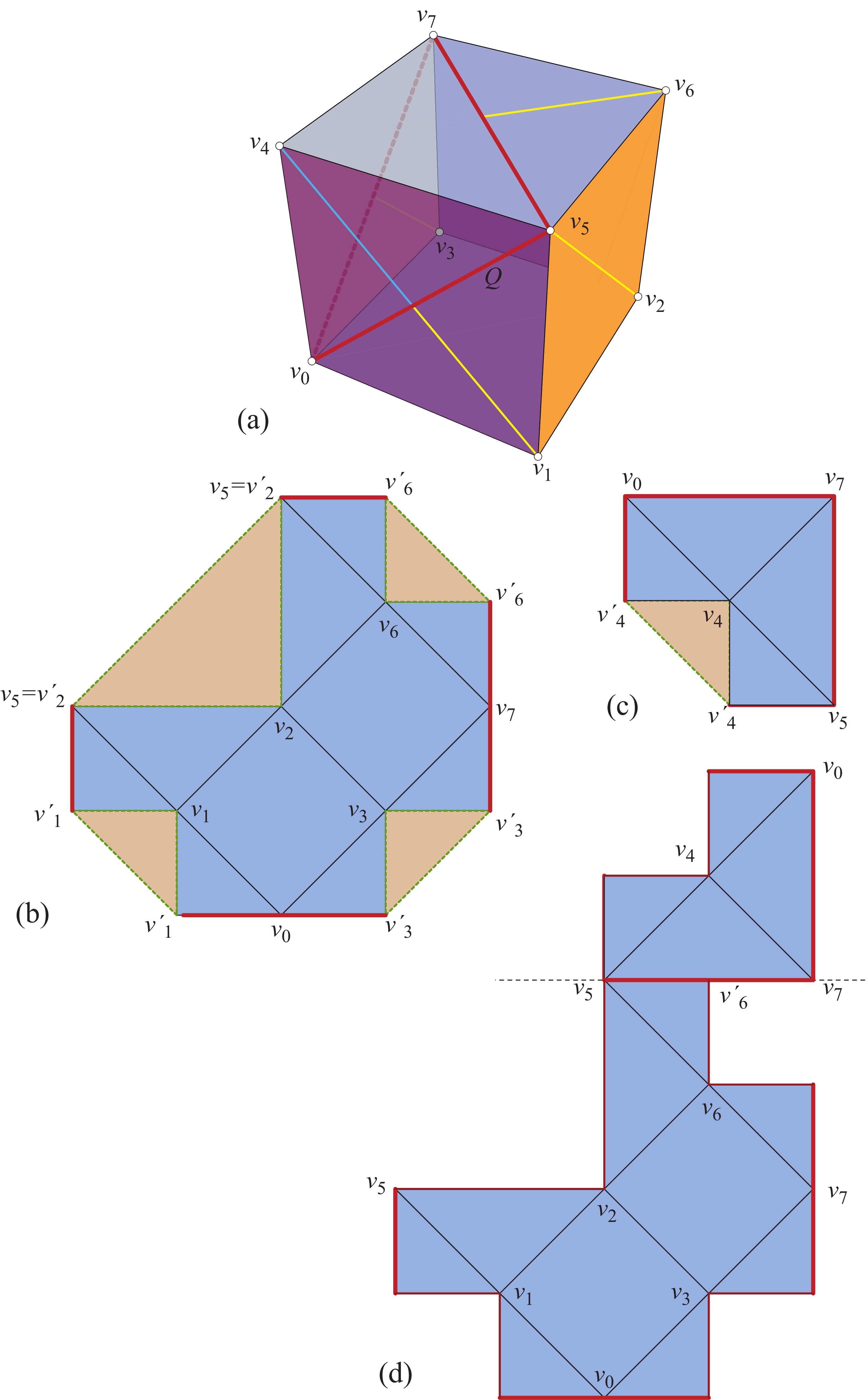}
\caption{
(a)~Cube and quasigeodesic $Q=(v_0,v_5,v_7)$. 
Shortest paths $\sp(v_i)$ as indicated.
(b)~Flattening the right half by insertion of curvature triangles
along the shortest paths $\sp(v_i)=v_iv'_i$.
(c)~Flattening the left half.
(d)~Two halves joined at $s=v_5 v'_6$.
}
\figlab{cube_geodesic_hemis}
\end{figure}

A central fact that enables our construction is this key lemma from
\cite[Cor.~1]{iiv-qfpcs-07}, slightly modified for our circumstances:%

\begin{lemma}  
Let $W$ be a simple closed polygonal curve on a convex surface $\P$, 
and let $p$ be any point of one (closed) half-surface $P$ bounded by $W$,
but not on $W$. Let $p'$ be one of the points of $W$ closest to $p$.
Then for any choice of $\sp(p)=pp'$,
the angle $\a$ made by $\sp(p)$ with $W$ at $p'$ is at least $\pi/2$.
In particular, if $p'$ is not a corner of $W$ 
then $\a=\pi/2$ and the path $\sp(p)$ is unique as shortest between $p$ and $p'$,
and $\a > \pi/2$ occurs only at corners $p'$ where the angle of $W$ toward $P$ 
is larger than $\pi$.                                                                 
\lemlab{IIV}                                                                    
\end{lemma}


A second fact we need concerning 
the shortest paths $\sp(v_i)$
is that they are disjoint.
\begin{lemma}
Any two shortest paths 
$\sp(v_1)$ and $\sp(v_2)$
are disjoint,
for distinct vertices $v_1,v_2 \in P$.
\lemlab{sp.disjoint}
\end{lemma}
\begin{proof}
Suppose for contradiction that at least one point $u$ is shared:
$u \in \sp(v_1)\cap \sp(v_2)$.
We consider four cases:
one shortest path is a subset of the other,
the shortest paths cross, the shortest paths touch at an interior point
but do not cross, or their endpoints coincide.
\begin{enumerate}
\item $\sp(v_2) \subset \sp(v_1)$.
Then $\sp(v_1)$ contains a vertex $v_2$ in its interior, which violates
a property of shortest paths~\cite[Lem.~4.1]{ss-spps-86}.
\item $\sp(v_1)$ and $\sp(v_2)$ cross properly at $u$.
It must be that $|uv'_1|=|uv'_2|$, otherwise both paths
would follow whichever tail is shorter.
But now it is possible to shortcut the path in the vicinity of $u$ via
$\s$ as shown in 
Figure~\figref{sp_lemma}(a), and the path $(v_1,\s,v'_2)$ is shorter than $\sp(v_1)$.
\item $\sp(v_1)$ and $\sp(v_2)$ touch at $u$ but do not cross properly there.
Then there is a shortcut $\s$ to one side (the side with angle less than $\pi$),
as shown in 
Figure~\figref{sp_lemma}(b).
\item $v'_1 = v'_2$.
From Lemma~\lemref{IIV} we know the two paths are orthogonal to the
quasigeodesic $Q$ at $v'_1 = v'_2$ hence, 
since they cannot be in the situation of Case~1, there is an angle $\a > 0$
separating the paths in a neighborhood
of the common endpoint; see Figure~\figref{sp_lemma}(c).
Then $Q$ has more than $\pi$ angle to one side at
this point, violating the definition of a quasigeodesic.
\end{enumerate}
\end{proof}

\begin{figure}[htbp]
\centering
\includegraphics[width=\linewidth]{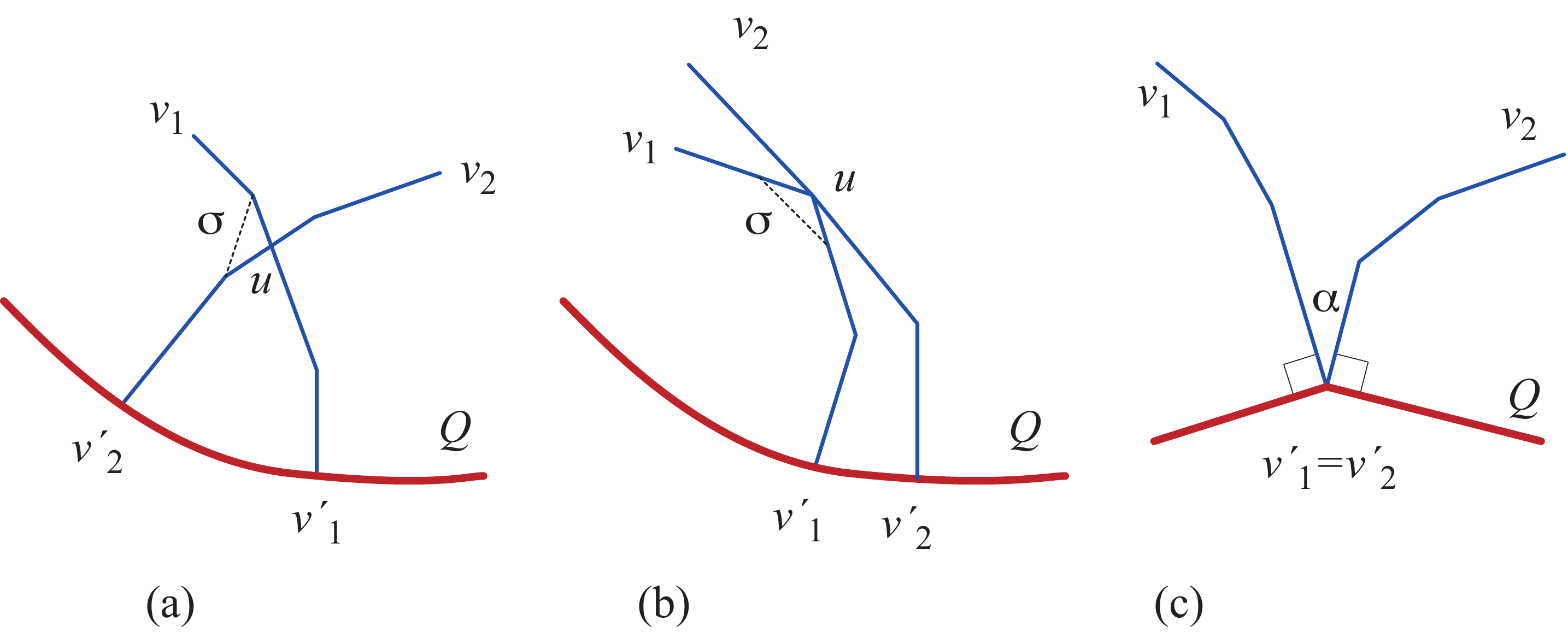}
\caption{Lemma~\protect\lemref{sp.disjoint}:
(a)~paths cross;
(b)~paths touch at an interior point;
(c)~paths meet at endpoint.}
\figlab{sp_lemma}
\end{figure}

This lemma ensures that the cuttings along $\sp(v_i)$ do not interfere
with one another.

\subsection{Flattening the Halves for Quasigeodesics}
The next step is to flatten each chosen
half $P_1$ and $P_2$ (independently) 
by suturing in ``curvature triangles''
along each $\sp(v)$ path.
Let $P$ be one of $P_1$ or $P_2$.
The basic idea goes back to Alexandrov~\cite[p.~241, Fig.~103]{a-cp-05},
and was used also in~\cite{iv-cfpcs-08}.
Let $\ell$ be the length of a shortest path $\sp(v)$,
and let $\o=\o(v)> 0 $ be the curvature at $v$.
We glue into $\sp(v)=v v'$ the isosceles \emph{curvature triangle}
$\triangle$ with apex angle $\o$ gluing
to $v$,
and incident sides of length $\ell$ gluing along the cut $v v'$. 
This is illustrated in
Figure~\figref{cube_geodesic_hemis}(b,c).
We display this in the plane for convenience of presentation;
the triangle insertion should be viewed as operations on the manifolds
$P_1$ and $P_2$, each independently.

This procedure only works if $\o < \pi$, for $\o$ becomes the apex of the
inserted triangle $\triangle$.
If $\o \ge \pi$, we glue in two triangles of apex angle $\o/2$, both with their
apexes at $v$.\footnote{
   One can view this as having two vertices with half
   the curvature collocated at $v$.
}
Slightly abusing notation, we use $\triangle$ to represent
these two triangles together.  
In fact we must have $\o < 2\pi$ for any vertex $v$
(else there would be no face angle at $v$),
so $\o/2 < \pi$ and this insertion is indeed well defined.

We should remark that an alternative method of handling $\o \ge \pi$
would be to simply not glue in anything to the 
vertex $v$ with $\o(v) \ge \pi$, in which
case we still obtain the lemma below 
leading to the exact same unfolding.

Now, because $\o$ is the curvature (angle deficit) at $v$, gluing in $\triangle$
there flattens $v$ to have total incident angle $2\pi$.
Thus $v$ is no longer a vertex of $P$ 
(and two new vertices are created along the bounding curve).

Let $P^\triangle$ be
the new manifold with boundary obtained after insertion of all 
curvature triangles into $P$.
We claim that a planar development of $P^\triangle$ does not
overlap; i.e., $P^\triangle$ is isometric to a simple planar polygonal domain. 

In the proof we
use two results of Alexandrov.
The first is his celebrated theorem~\cite[Thm.~1, p.~100]{a-cp-05}
that gluing polygons to form a topological sphere in such a way that at
most $2\pi$ angle is glued at any point, results in a unique
convex polyhedron.

The second is a tool that relies on this theorem:
\begin{lemma}
Let $M$ be a polyhedral manifold with convex boundary, i.e., 
the angle towards $M$ of the left and right tangent directions
to $\bM$ is at most $\pi$ at every point.
Then the closed manifold $M^\#$ obtained by gluing two copies of
$M$ along $\bM$ is isometric to a unique polyhedron, with
a plane of symmetry $\Pi$ containing $\bM$, which is a convex polygon in $\Pi$.
\lemlab{Doubling}
\end{lemma}
\begin{proof}
That $M^\#$ is a convex polyhedron follows from 
Alexandrov's gluing theorem.
Because $M^\#$ has intrinsic symmetry, 
a lemma of
Alexandrov~\cite[p.~214]{a-cp-05} applies to show that the
polyhedron has a symmetry plane containing the polygon $\bM$.
\end{proof}

Now we apply these lemmas to $P^\triangle$:

\begin{lemma}
For each half $P$ of $\P$,
$P^\triangle$ is a planar convex polygon,
and therefore simple (non-overlapping).
\lemlab{convex.polygon}
\end{lemma}
\begin{proof}
$P^\triangle$ is clearly a topological disk: $P$ is, 
and the insertions of $\triangle$'s maintains it a disk.
At every interior point of $P^\triangle$, the curvature is zero by construction.
So the interior is flat.

\hide{
Let $\o_Q$ be the total curvature enclosed within $Q$ on $P$,
and $\tau_Q$ the total turn of $Q$, i.e., the turn of $\bP$.
The Gauss-Bonnet Theorem yields $\tau_Q + \o_Q = 2\pi$.
This is precisely the total turn of $\bP^\triangle$, because
that boundary turns $\tau_Q$, plus a total of $\o_Q$ for all the
inserted curvature triangles.
So indeed the boundary of $P^\triangle$ turns just as much as it should
if it is a planar polygon.  It remains to establish
that it is a simple polygon.
}

Next we show that the boundary 
$C=\bP^\triangle$
is convex.
This follows from the orthogonality of $\sp(v)$ guaranteed by Lemma~\lemref{IIV},
as the base angle of the inserted triangle(s) is
$\pi/2 - \o/2$ for $\o < \pi$,
or $\pi/2 - \o/4$ for $\o \ge \pi$
(see Figure~\figref{Q_convex}; $\o=\o(v)$),
so the new angle is smaller than $\pi$ by $\o/2$ or $\o/4$.
\begin{figure}[htbp]
\centering
\includegraphics[width=0.75\linewidth]{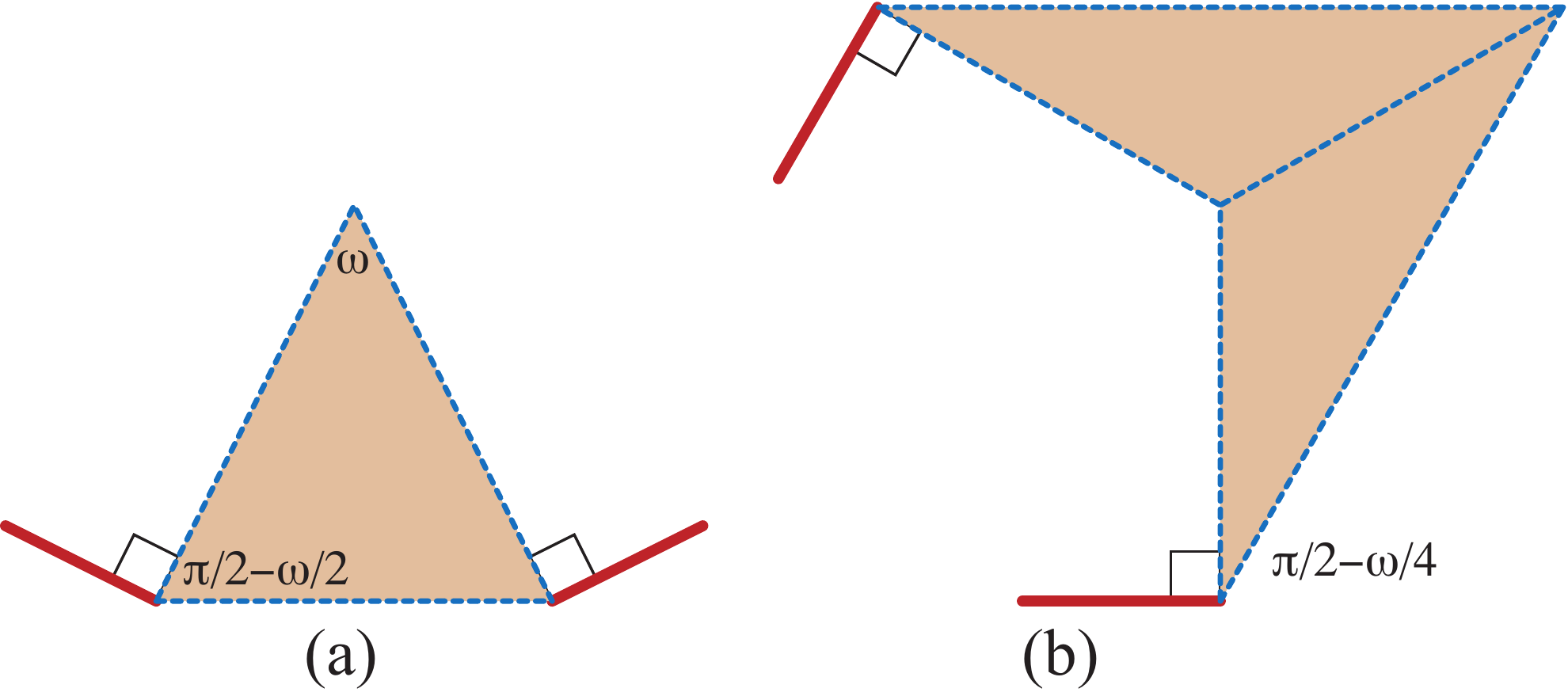}
\caption{Lemma~\protect\lemref{convex.polygon}, Case~1: 
(a)~$\o < \pi$; (b)~$\o \ge \pi$.}
\figlab{Q_convex}
\end{figure}

Now we know that $P^\triangle$ is homeomorphic to a disk, and it has
a convex boundary $C$.  
We next prove it is isometric to a planar convex polygon
by applying Lemma~\lemref{Doubling}.
Let $P^\#$ be the result of gluing
two copies of $P^\triangle$ back-to-back along $C$.
The lemma says that $P^\#$ has a symmetry plane
containing $C$.
As all the vertices of $P^\#$ are on $C$, $P^\#$ 
itself must be planar (and doubly covered).
Therefore $P^\triangle$ is a planar convex polygon, and therefore simple.
\end{proof}

Note that, when the total curvature in $P$ is $2\pi$
then the straight development
of $Q$ is turned $2\pi$ by the $\triangle$ insertions,
as in Figure~\figref{cube_geodesic_hemis}(b).
When the total curvature in $P$ is less than $2\pi$,
the development of $Q$ is not straight, but the $\triangle$ insertions
turn it exactly the additional amount needed to close it to $2\pi$,
as in~(c) of the same figure.

\subsection{Joining the Halves for Quasigeodesics}
The third and final step of the unfolding procedure
selects a \emph{supporting segment} $s \subset Q$
whose relative interior does not contain
a projection $v'$ of a vertex.
All of $Q$ will be cut except for $s$.
When $Q$ is a closed quasigeodesic,
any choice for $s$ generates a supporting line to a planar development of $P_i$,
$i=1,2$,
because $P_i^\triangle$ is a convex domain.
Then joining planar developments of $P_1$ and $P_2$ along $s$
places them on opposite sides of the line through $s$,
thus guaranteeing non-overlap.
See Figure~\figref{cube_geodesic_hemis}(d),
where $s=v_5 v'_6$.

\section{Quasigeodesic Loops}
We now turn to the case when $Q$ is a closed quasigeodesic loop
with loop point $x$, at which the angle toward $P$ is $\b > \pi$.
See Figure~\figref{GeodesicLoopCube}, where $\b = \frac{3}{2} \pi$.

\subsection{Shortest Path Cuts for Quasigeodesic Loops}
Lemma~\lemref{IIV}, claiming that the shortest path
$\sp(p)=p p'$ is unique and orthogonal to $Q$ at $p'$, needs no
modification.
For $p'=x$, the shortest path $p x$ makes a non-acute ($\ge \pi/2$)
angle with $Q$ on both sides.
Lemma~\lemref{sp.disjoint}, claiming disjointness of the
shortest paths for distinct vertices $v_1,v_2$,
need only be modified by observing that more than one shortest
path can project to $x$, but the paths do not intersect other than at $x$.
The proof is identical.

The significant differences between quasigeodesics and quasigeodesic loops
are concentrated largely in the proof for flattening,
and in the proof for joining the halves.
We should emphasize that the three algorithm steps detailed
earlier are the same;
it is only the justification that becomes more complicated.

\subsection{Flattening the Halves for Quasigeodesic Loops}
We aim to prove the analog of Lemma~\lemref{convex.polygon}
for quasigeodesic loops: $P^\triangle$ is (isometric to) a simple, planar
polygon.
Obviously that lemma establishes this for the convex half
of $\P$.
Let $P$ be the nonconvex half, in which the angle $\b$
at $x$ exceeds $\pi$.

The argument will be different (and easier) if no vertices project to $x$.

\subsubsection{No Vertex Projects to $x$}

\begin{lemma}
If no vertex of $P$ projects to $x$, 
then $P^\triangle$ is a simple, planar polygon.
\lemlab{NoVertex}
\end{lemma}
\begin{proof}
Let $\g$ be a geodesic on $P^\triangle$ bisecting the angle $\b$ at $x$,
and meeting $\bP^\triangle$ at $y$;
see Figure~\figref{NoVertex}.
\begin{figure}[htbp]
\centering
\includegraphics[width=0.5\linewidth]{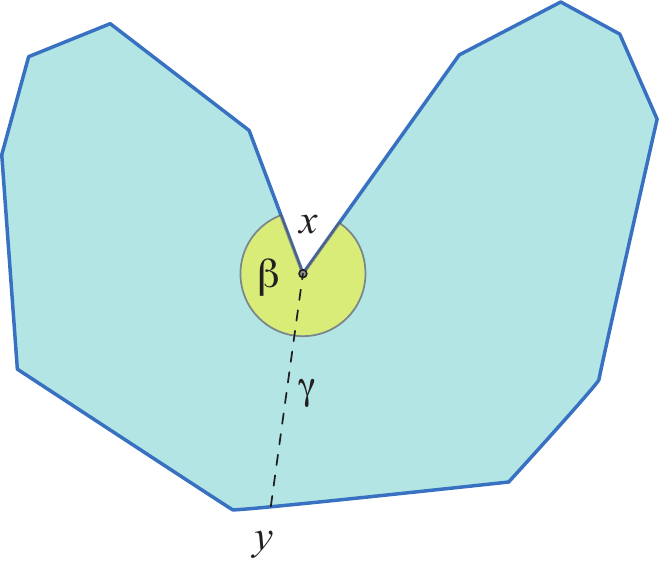}
\caption{No vertex of $P$ is incident to $x$.}
\figlab{NoVertex}
\end{figure}
Because $P^\triangle$ is flat, $\g$ does in fact meet the boundary
of $P^\triangle$, i.e., it does not self-cross before reaching the boundary.
Cut $P^\triangle$ along $\g=xy$ into two manifolds.
Because $\b < 2\pi$, the angle at $x$ in each manifold
is less than $\pi$.
Therefore, we have obtained exactly the situation
in
Lemma~\lemref{convex.polygon}: each is a flat manifold homeomorphic
to a disk, with a convex boundary.
Each is therefore isometric to a planar convex polygon,
with a supporting line through $xy$.
Therefore they may rejoin along $\g$ to form a planar simple polygon,
with one reflex (concave) vertex at $x$.
\end{proof}

\subsubsection{Vertices project to $x$}
Now we consider the more difficult case, when 
vertices in a set
$V=\{v_1,v_2,\ldots,v_k\}$ project to $x$.
Rather than construct $P^\triangle$ directly as above,
we proceed in two stages:
first we insert curvature triangles for all those vertices
of $P$ that project to $Q \setminus \{x\}$,
forming a (nonflat) manifold $X$, and then argue that
$X$ unfolds without overlap.  We will not insert curvature triangles
for the vertices in $V$, 
but rather argue directly for non-overlap
of a flattening $U(X)$ of $X$.
The boundary of $U(X)$ will consist of two portions: a convex chain $C$
deriving from $Q \setminus \{x\}$ and the insertion of the curvature triangles,
and a nonconvex chain $N$ deriving from $x$ and the vertices $V$.
Much as in Figure~\figref{NoVertex}, we will cut $X$ by a geodesic $\g$
and later glue the two unfolding pieces back along $\g$.
We will see that the chain $N$ is formed of subchains of particular star unfoldings
with respect to $x$.

\paragraph{Preliminaries.}
We need several ingredients for this proof.
First, we will need the cut locus of $x$ on the manifold-with-boundary $X$,
which may be defined as in Section~\secref{Introduction}:
$C(x)$ is
the closure of the set of all points of $X$ joined to the source
point $x$ by at least two        
shortest paths on $X$.

Second, we need two extensions to the
theorem
established in~\cite[Thm.~9.1]{ao-nsu-92} that the star unfolding
of a polyhedron does not overlap.
That paper assumed the source point $x$ was generic in the sense that
it had a unique path to every vertex.
We extend the result to the case 
when $x$ has two or more distinct shortest
paths to a vertex $v$: then simply cutting one of the paths to $v$
again leads to non-overlap.
The second extension is more substantive.
For any three points $r,s,t$ in the plane, forming the counterclockwise angle
$\t = \angle rst$ at $s$, define the \emph{wedge} determined by
these points to be the set of points $p$ such that the segment $ps$
falls within the angle $\t$: counterclockwise of $rs$ and clockwise of $ts$.
So the wedge is an angle apexed at $s$ when $\t < \pi$, and the
complement of an angle when $\t > \pi$.
\begin{lemma}
Let $v$ be a vertex of the star unfolding $S(x)$ 
of a polyhedral convex surface $\P$
with respect to $x \in \P$,
and $x_1$ and $x_2$ the images of $x$ resulting from cutting a shortest path $xv$.
Then $S(x)$ is disjoint from the open wedge determined by
$x_1,v,x_2$, or, equivalently, $S(x)$ is enclosed in the closed wedge
determined by $x_2,v,x_1$.
\lemlab{wedge}
\end{lemma}
See Figure~\figref{StarWedge}.
\begin{figure}[htbp]
\centering
\includegraphics[width=0.75\linewidth]{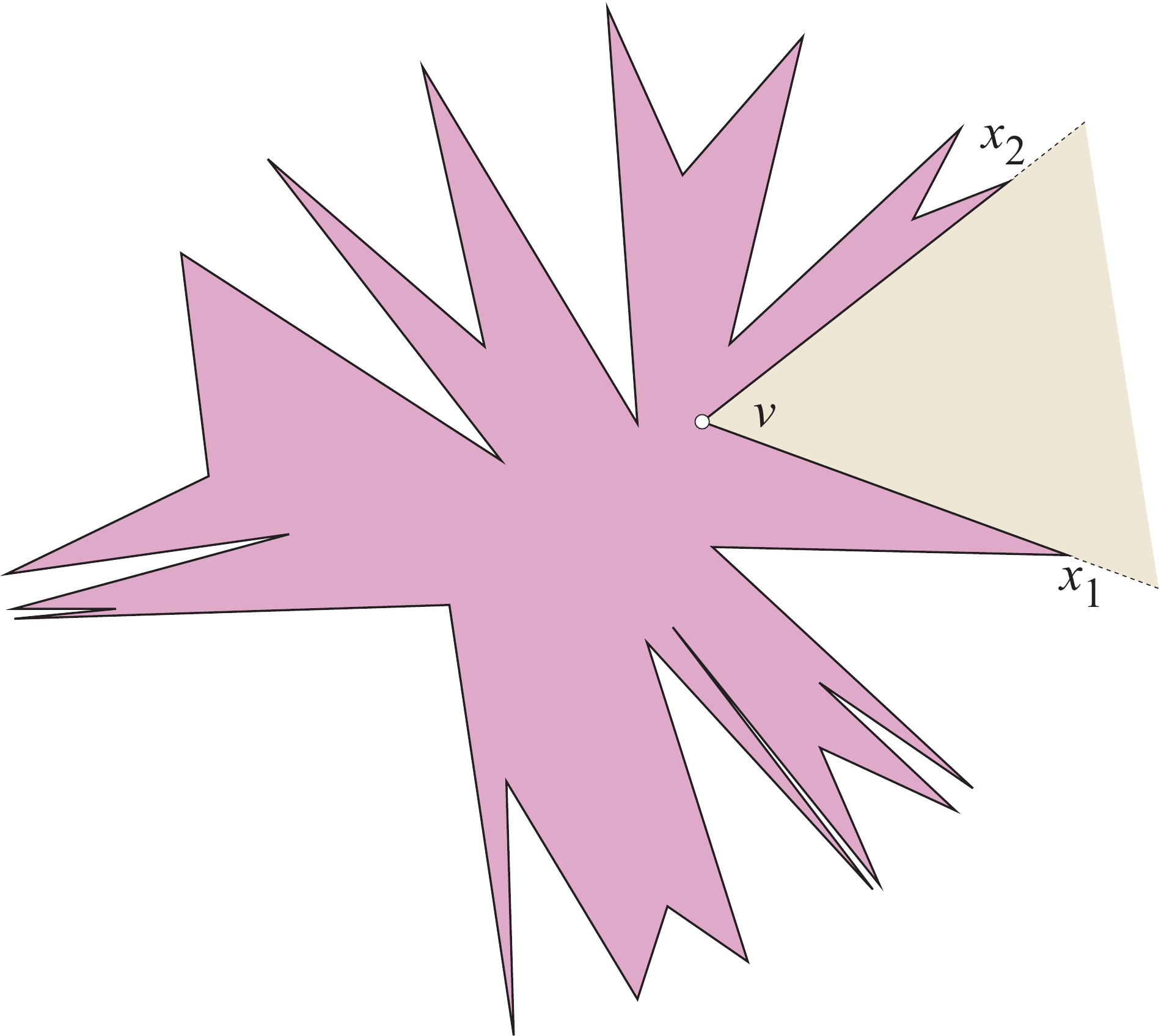}
\caption{The open wedge 
determined by 
$x$ and the vertex $v$
is disjoint from the star unfolding $S(x)$.
}
\figlab{StarWedge}
\end{figure}
Both of these extensions to~\cite{ao-nsu-92}
are established in~\cite{ov-npsuc-09}.


We will also need this fact:
\begin{lemma}
Let $M^\#$ be a convex polyhedron obtained by doubling $M$
as in Lemma~\lemref{Doubling}.
Then a shortest path $vu$ from a vertex
$v$ on the symmetry plane $\Pi$, to a vertex $u$ not on $\Pi$,
lies completely in the half containing $u$.
\lemlab{Msp}
\end{lemma}
\begin{proof}
Let $u$ be in the bottom half.
Suppose for contradiction that a shortest path $vu$ includes a section $\sigma$
that enters the top half at $s_1 \in \Pi$ and exits back to the bottom
half at $s_2 \in \Pi$. Then we can reflect $\sigma$ through $\Pi$ to
place it on the bottom half, without changing its length.
Because shortest paths on convex surfaces do not branch, we obtain a contradiction.
\end{proof}

The flattening proof consists of two parts: first, a procedure is used to
find the splitting geodesic $\g$, or to determine no such geodesic
exists.  The second part then has two cases for reaching the
final unfolding.

\paragraph{Part~I: Excising Digons.}
We now describe a process to flatten a subset of $X$ by cutting out 
regions containing all its
interior vertices.  The flattened version $X'$ will then be used
in further steps of the proof.

Recall we defined $X$ as the manifold obtained by inserting curvature triangles
into $P$
for all shortest paths incident to $Q \setminus \{x\}$,
but leaving $V=\{v_1,v_2,\ldots,v_k\}$ unaltered.
So $X$ is not flat, but its boundary $\bX$ is the 
simple closed convex curve $C$.

Let $C(x)$ be the cut locus of $x$ on $X$.
In general $C(x)$ is a forest of trees, each of which
meets $\bX$ in a single point, and whose leaves are the vertices
in $V$.
This follows from the proof of Theorem~A in~\cite{st-cldsa-96},
but may also be seen intuitively as follows.
A remark of Alexandrov~\cite[p.~236]{a-cp-05}
shows that $X$ may be extended to a convex polyhedral
surface $\X$.  The cut locus of $x$ on $\X$ is a tree,
which gets clipped by $\bX$ to a forest on $X$.
Another way to see this is to observe that,
at every vertex $v$ interior to $X$, a geodesic edge of $C(x)$
emanates, and when two such edges meet interior to $X$, they join
to start a third edge of $C(x)$. 

Let $w_1,w_2,\ldots$ be the points of $C(x) \cap \bX$.
Each $w_i$ is joined to $x$ by at least two shortest paths, and
the union of two of these bounds a 
digon $D_i$ that includes the tree component $T_i$ of $C(x)$ incident to $w_i$. 
                                                                                
Cut out from $X$ all $D_i$'s, and glue back their boundary segments in 
$X \setminus \bigcup_i {\mathrm int} D_i$
to get a new surface $X'$.
Because the vertices $V$ 
of $X$ have been excised with the digons,
$X'$ is flat.
And $\bX'$ is convex except (possibly) at $x$.
Therefore, the argument of Lemma~\lemref{NoVertex} applies to $X'$,
showing that it is the join of two planar convex polygons.

We illustrate this construction before proceeding with the proof,
first with a real example, second with an abstract example.
Figure~\figref{DigonCube}(a) illustrates $X$ for the geodesic loop
on a cube from Figure~\figref{GeodesicLoopCube}; only
$v_1$ projects to $x$.
$C(x)$ is in this case the single segment $x w_1$.
After excising the digon bound by the two $x w_1$ shortest paths,
the planar polygon $X'$ shown in~(b) is attained.
\begin{figure}[htbp]
\centering
\includegraphics[width=0.9\linewidth]{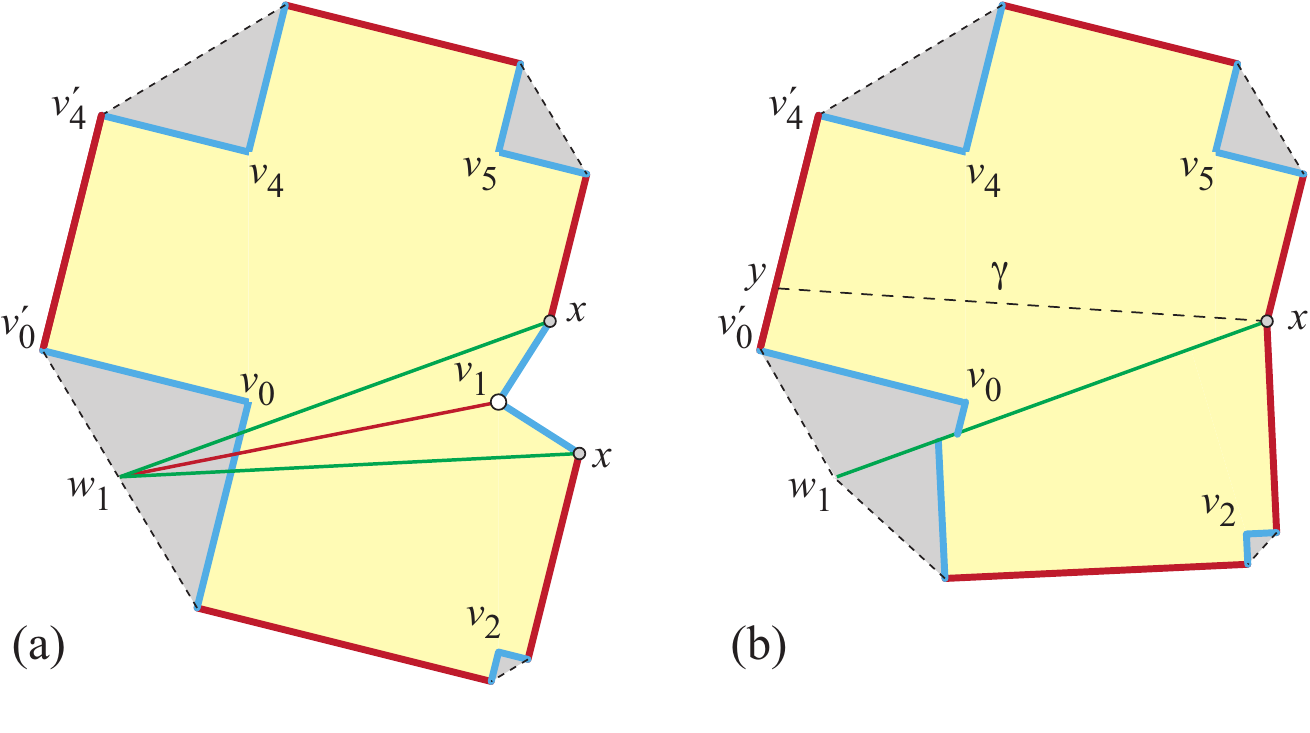}
\caption{(a)~Manifold $X$ for the nonconvex half
of Figure~\protect\figref{GeodesicLoopCube}(b).
The two points labeled $x$ should be identified
along the shortest path $v_1 x$; 
so $X$ in this case is isometric to a subset of a cone with apex $v_1$.
(b)~After removing the digon bounded by paths from $x$ to $w_1$
and identifying the paths.}
\figlab{DigonCube}
\end{figure}

A more generic situation is illustrated in Figure~\figref{DigonsAbstract}.
\begin{figure}[htbp]
\centering
\includegraphics[width=\linewidth]{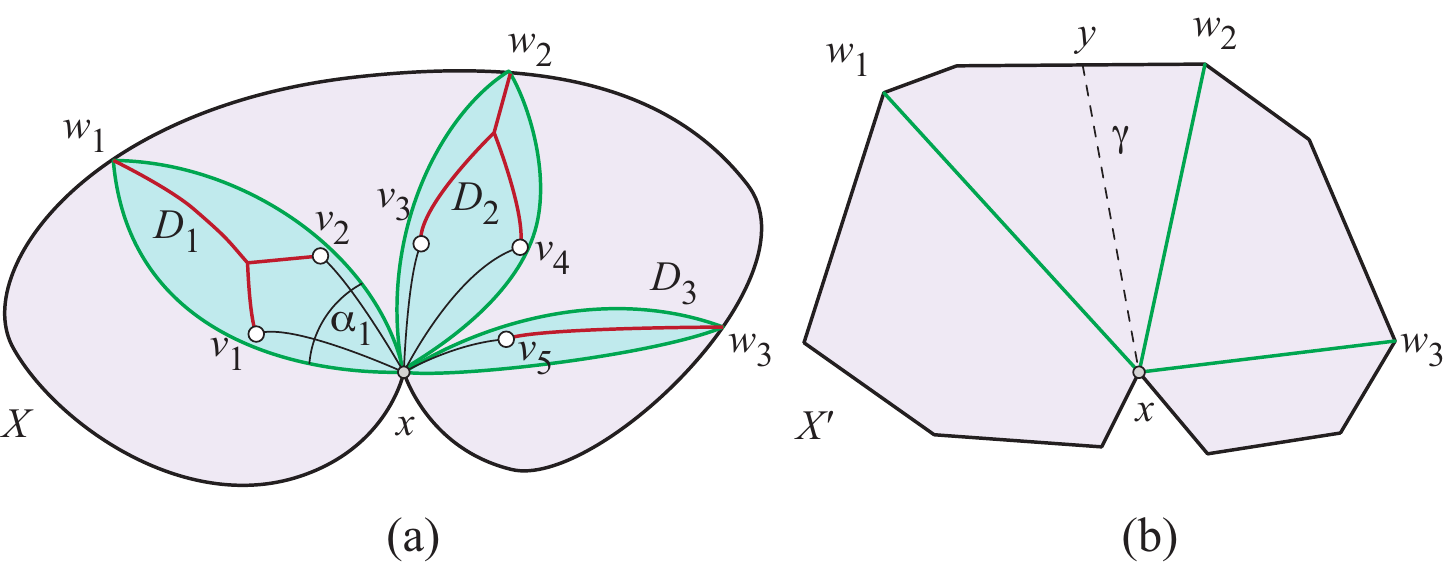}
\caption{(a)~The manifold $X$ depicted abstractly.
(b)~Planar, simple $X'$ after excision of digons.
}
\figlab{DigonsAbstract}
\end{figure}

\paragraph{Part~II: Partitioning $X$, Case~1.}
The next step of the proof is to partition the manifold $X$ into pieces
each of which has a convex boundary.
The partition will be into either two or three pieces.
We start with the two-piece Case~1.
Let $\a_i$ be the digon angle of $D_i$ at $x$ on $X$.
Then the angle $\b$ at $x$ on $X$ is reduced by $\sum \a_i$ on $X'$.
Case~1 holds when there is a segment $\g=xy$ on $X'$ from $x$ to 
some point $y \in \bX'$
such that the corresponding geodesic on $X$ splits the angle $\b$ at $x$
into two parts, each of which is at most $\pi$.
(It is not possible to simply select the bisector of $\b$ on $X$
for $\g$, because that bisector might fall inside the $\a_i$ of one
of the digons, and so not be realized in $X'$.) 
Assuming for Case~1 that such a $\g$ exists, we partition $X$ by cutting along $\g$,
resulting in two (nonflat) manifolds $X_1$ and $X_2$.
Note that each such manifold has a convex boundary: 
$\bX$ is convex at every point excepting $x$, 
the angle at $y$ is less
than $\pi$, and we have conveniently split the angle at $x$.
Therefore, we may apply the ``doubling'' 
Lemma~\lemref{Doubling} to each half.
Henceforth we fix our attention to $X_1$.

According to that lemma, we have a convex polyhedron $X_1^\#$ with $\bX_1$
lying in the symmetry plane $\Pi$.
We'll call the portion above and below $\Pi$ the upper and lower halves respectively.
The upper half includes
(a subset of) vertices of $V$, say $v_1,\ldots,v_j$, 
and the lower half includes equivalent copies, call them $u_1,\ldots,u_j$.
There are a number of vertices on $\Pi$ deriving from $\bX_1$,
including $x$, $y$; call them $z_0, z_1, z_2, \ldots$,
with $z_0$ adjacent to $y$.

\begin{figure}[htbp]
\centering
\includegraphics[width=\linewidth]{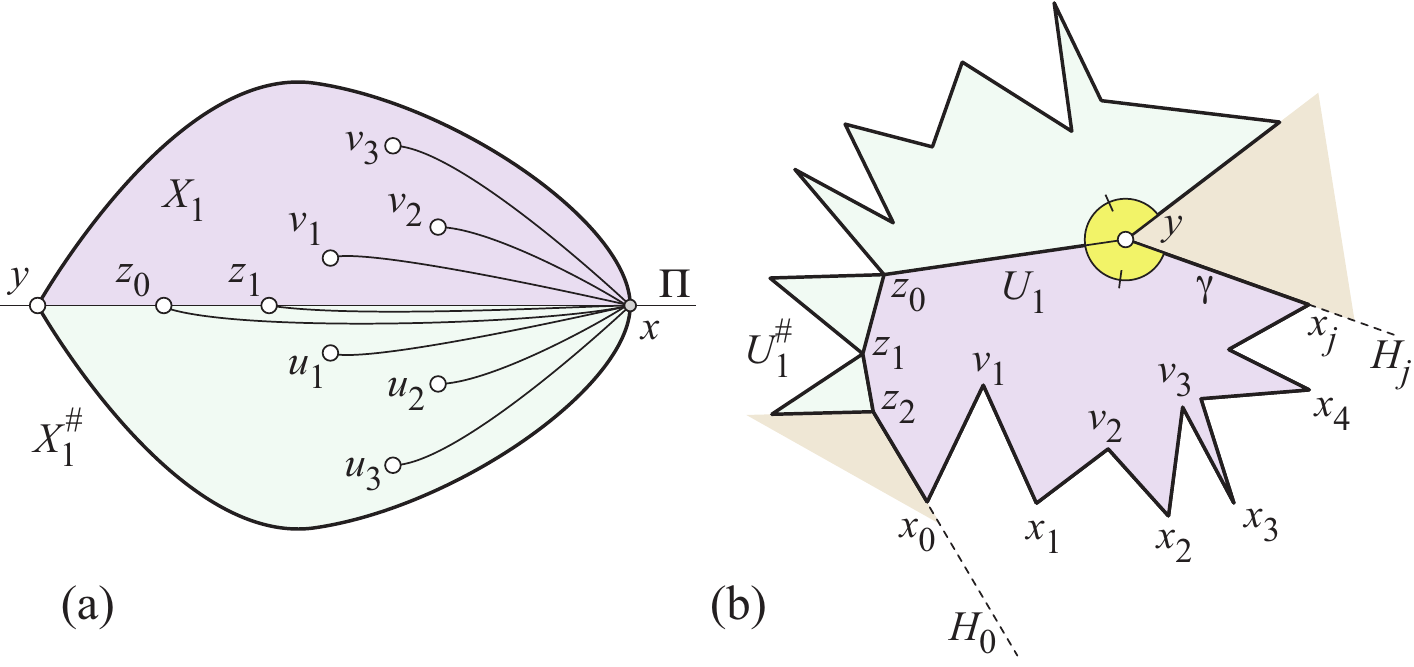}
\caption{(a)~The double polyhedron $X_1^\#$ of $X_1$.
(b)~The star unfolding $U_1^\#$ of $X_1^\#$ with respect to $x$.}
\figlab{X1hash}
\end{figure}
Now the plan is to construct the star unfolding of $X_1^\#$ with respect
to the point $x$.
By Lemma~\lemref{Msp}, the shortest paths from $x$ to the vertices 
of $X^\#_1$ in
$\Pi$ 
lie wholly in one half or the other.
For a vertex on $\Pi$, either the shortest path lies in $\Pi$,
or there are pairs of equal-length shortest paths, one in the upper
and one in the lower half.
By our extension of the star unfolding theorem, we may choose
which shortest path to cut in the case of ties, and still obtain a non-overlapping
unfolding.
We choose to select all shortest paths to the vertices on $\Pi$ in the lower
half (or along $\Pi$ if that is where they lie).
See Figure~\figref{X1hash}(a).
Note that $\g=xy$ will necessarily lie in $\Pi$, and it is also
necessarily a shortest path from $x$ to $y$
(because
it is an edge of the polyhedron, and every edge of a polyhedron is a shortest
path between its endpoints).  So we cut $xy=\g$ as well.
The result is a planar, non-overlapping unfolding $U_1^\#$.
Now, identify within $U_1^\#$ the portion $U_1$ corresponding
to the upper half.
Also note that $\g$ is one edge of $\bU_1$.
Let $x_{i-1}$ and $x_i$ be the images of $x$ adjacent to $v_i$. 
See Figure~\figref{X1hash}(b).
Note that $N_1=(x_0,v_1,x_1,v_2,\ldots,v_j,x_j) \subset \bU_1$ is a subchain of
the nonconvex chain $N$ mentioned earlier, and its complementary
chain $C_1$ in $\bU_1$ is a subchain of the convex chain $C$.

Next we perform the exact same procedure for $X_2$, resulting in $U_2^\#$
containing $U_2$.
Now glue $U_1^\#$ to $U_2^\#$ along their boundary edges deriving from $\g$.
Applying Lemma~\lemref{wedge} at the point $y$ in each unfolding
shows that they join without overlap, as follows.
\begin{figure}[htbp]
\centering
\includegraphics[width=0.75\linewidth]{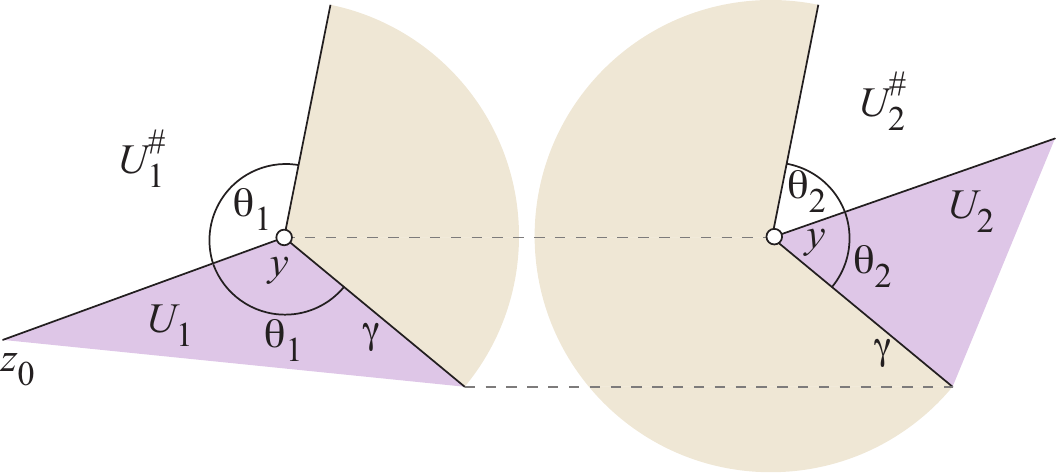}
\caption{Complementary exterior open wedges at $y$.}
\figlab{CompWedges}
\end{figure}
Let $\t_1$ be the angle at $y$ in $U_1$, and $\t_2$ the angle in $U_2$.
If $y$ coincides with a vertex of $Q$, then $\t_1 + \t_2 < \pi$,
hence joining the closed wedges enclosing $U_1^\#$ and $U_2^\#$ 
leaves at $y$ an empty open wedge of measure
$2( \pi - \t_1 - \t_2 )$.
If $y$ is not a vertex of $Q$, then $\t_1 + \t_2 = \pi$,
hence the closed wedges enclosing $U_1^\#$ and $U_2^\#$ are complementary;
see Figure~\figref{CompWedges}.
Thus, in either case, the contained $U_1$ and $U_2$ do not overlap one another.
And so we have established that $U(X) = U_1 \cup U_2$ is
a planar simple polygon.

\paragraph{Part~II: Partitioning $X$, Case~2.}
Case~1 relies on the existence of a segment $\g$ on $X'$ 
such that the corresponding geodesic on $X$ splits the angle $\b$ at $x$
into two parts, each of which is at most $\pi$.
Now we consider the possibility that there is no such $\g$.
We illustrate this possibility with an example before handling this case.

In Figure~\figref{Case2b},
$\P$ is a tetrahedron with three right angles incident to $v_0$.
The geodesic loop shown has at $x$ an angle 
$\b=330^\circ$ in the lower half,
and two vertices $v_1$ and $v_2$, included
in the same digon, projecting to $x$.
In this example, the sole digon constitutes the majority of $X$,
\begin{figure}[htbp]
\centering
\includegraphics[width=0.9\linewidth]{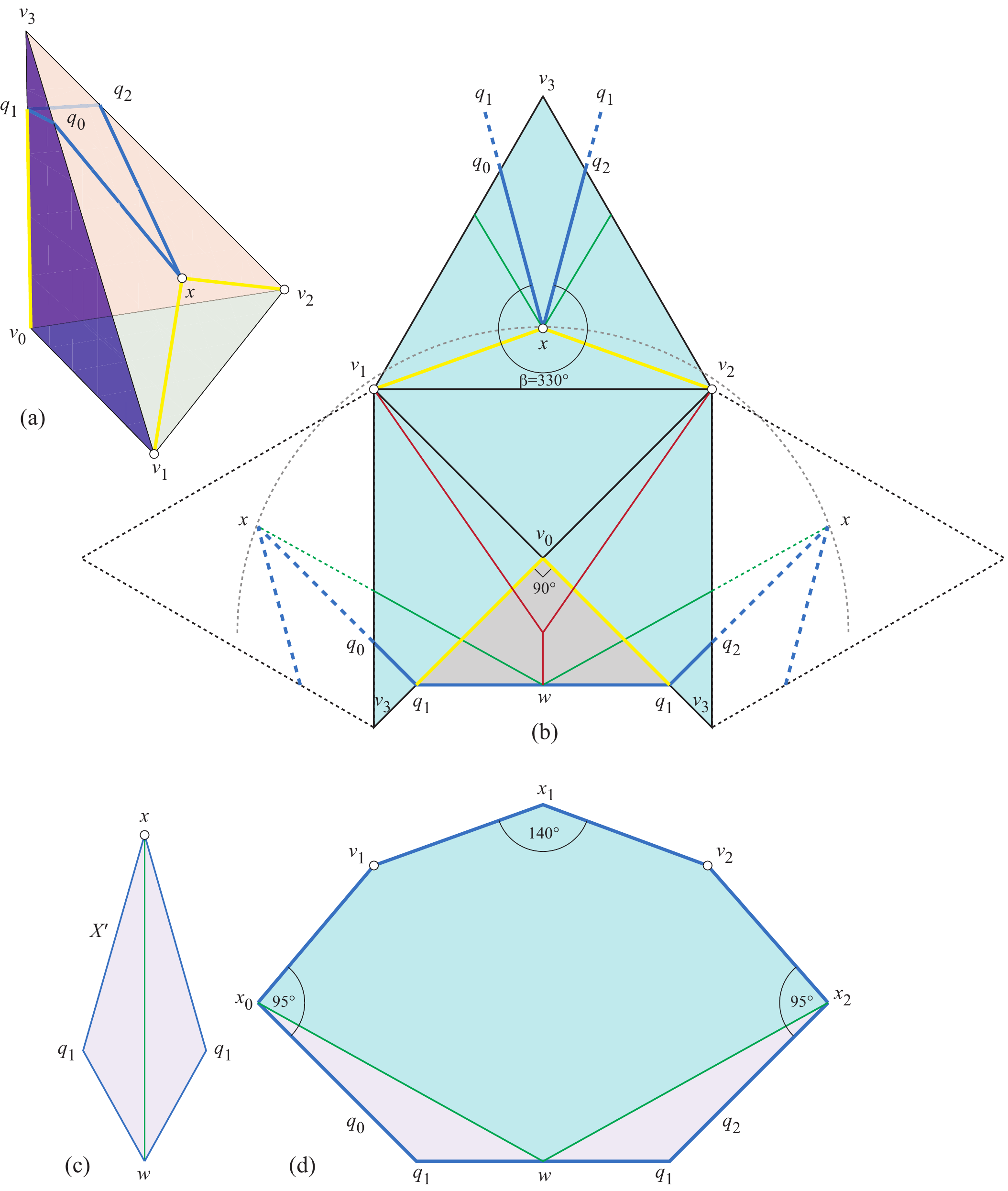}
\caption{(a)~Geodesic loop $Q=(q_0,q_1,q_2,x)$.
(b)~The manifold after insertion
of the curvature triangle
with $\o(v_0)=90^\circ$ at $v_0$.
$C(x)$ is the `Y', and the digon is
bounded by $xw$.
(c)~$X'$: after removal of digon.
(d)~$U(X)$. Note the angles at the three images of $x$ sum
to $\b=330^\circ$.}
\figlab{Case2b}
\end{figure}
and its removal leaves $X'$ ((c) of the figure)
so narrow as to not admit a $\g$ with the
desired angle-splitting properties.

In this case, we have a ``fat'' digon $D_i$ (possibly $D_i=X$)
whose angle $\a_i$ at $x$
covers all the possible splitting segments $\g$.
Let $D=D_i$ to ease notation, and
let $\g_1$ and $\g_2$ be the two boundary edges (shortest paths) 
of $D$ connecting $x$ to
$w_i$.
Let $a$ and $b$ be the vertices adjacent to $x$ on $\bX$.
The angle from $ax$ to $\g_1$,
and the angle from $bx$ to $\g_2$, are both less than $\pi$.
Now we define three (topologically) closed submanifolds of $X$:
$X_1$, the portion bounded by $\g_1$ and containing vertex $a$,
$X_2$, the portion bounded by $\g_2$ and containing vertex $b$,
and the digon $D$ in between.
$X_1$ and $X_2$ have convex boundaries, just as the manifolds in Case~1,
and we go through the identical process:
$$X_i \rightarrow X_i^\# \rightarrow U_i^\# \rightarrow U_i \;.$$
The digon $D$ might not be convex at $x$;
in Figure~\figref{Case2b}(a) the angle at $x$ is $\a_i=300^\circ$.
So we cannot use the doubling lemma.
Instead we glue (``zip'') $\g_1$ to $\g_2$, which produces a convex polyhedron
$D^z$ containing the vertices of $V$ inside $D$,
and vertices at $x$ and at $w_i$.
Now we produce the star unfolding of $D^z$ with respect to $x$.
Note that $x w_i = \g_1 = \g_2$ is a shortest path on $D^z$,
so that gets cut as part of the star unfolding.
Call this unfolding $U_{D}$.
By Lemma~\lemref{wedge}, $U_{D}$ is contained within the closed wedge
at $w_i$, with wedge rays along $\g_1$ and $\g_2$.

Finally, we join $U_1^\#$ to $U_{D}$ along $\g_1$, 
and $U_2^\#$ to $U_{D}$ along $\g_2$,
nonoverlapping
by the wedge properties,
and conclude that the contained $U(X) = U_1 \cup U_{D} \cup U_2$
is a planar simple polygon.
See Figure~\figref{Case2b}(d).

Finally, removing the inserted curvature triangles from $U(X)$
establishes the analog of Lemma~\lemref{convex.polygon}:

\begin{lemma}
For $Q$ a quasigeodesic loop,
the star unfolding of the nonconvex half $P$ of $\P$
is a planar, simple polygon.
\lemlab{qloop}
\end{lemma}

\subsection{Joining the Halves for Quasigeodesic Loops}
Let $s$ be the edge of the unfolding $U=U(X)$ on which $y$ lies
(in Case~1) or on which $w$ lies (in Case~2).
If $y$ or $w$ is at a vertex of $\P$, then let $s$ be either incident
edge.  We claim that the line $L$ containing $s$ is a supporting
line to $U(X)$.
We establish this by identifying a larger class of supporting lines, 
which includes $L$.
We make the argument for Part~II, Case~1 above (when $\g$ exists),
as Case~2 is very similar.

Returning to Figure~\figref{X1hash}(b),
consider the two empty wedges exterior to $U_1^\#$, incident
to the vertex adjacent to $x_0$ that is not $v_1$
($z_2$ in that figure),
and the vertex
adjacent to $x_j$ that is not $v_j$, i.e., $y$.
Let $H_0$ and $H_j$ be the halflines on those wedge boundaries
that include $x_0$ and $x_j$ respectively.
The empty wedges imply that $U_1$ is included in the convex region $\overline U_1$
of the plane delimited by $H_0$, $H_j$, and the convex 
boundary $C_1 \subset \bU_1$:
the nonconvex boundary portion $N_1 \subset \bU_1$ can
cross neither $H_0$ nor $H_j$.
Let $H_0 \cap H_j$ be the point $p_1$, which might be ``at infinity''
if those halflines diverge.
Similarly, $U_2$ is included in a region $\overline U_2$
delimited by $H_j$ and $H_k$, which meet at $p_2$ or not at all
if those halflines diverge.
Let $L_j$ be the line containing $H_j$, which contains $\g$,
and $p_1$ and $p_2$ when those exist.
We consider three cases.

\begin{enumerate}
\item \emph{Both $p_1$ and $p_2$ exist.}
See Figure~\figref{Joining}(a).
Then there are two supporting lines parallel to $L_j$, 
and every edge between them along $\bU$ is supporting to $\overline U$ and so to $U$.
\item \emph{$p_2$ exists, but $p_1$ does not.}
See Figure~\figref{Joining}(b).
We have the one line parallel to $L_j$ supporting $\overline U_2$.
Let $p'$ be the point on $L_j$ on the opposite side of
$y$
from $p_2$ that maximizes the chain of $\partial \overline U_1$ visible from $p'$.
Then again every edge between the two supporting lines is 
supporting to $\overline U$ and so to $U$.
\item \emph{Neither $p_1$ nor $p_2$ exist.}
See Figure~\figref{Joining}(c).
Then again let $p'$ be the point on $L_j$ on the opposite side of $y$
that maximizes the visibility of 
$\partial \overline U$.
Then all the edges between the tangency points extend to supporting line.
\end{enumerate}
In particular, we see that the line(s) extending edge(s) incident to $y$ 
are among the supporting lines, as claimed.
\begin{figure}[htbp]
\centering
\includegraphics[width=\linewidth]{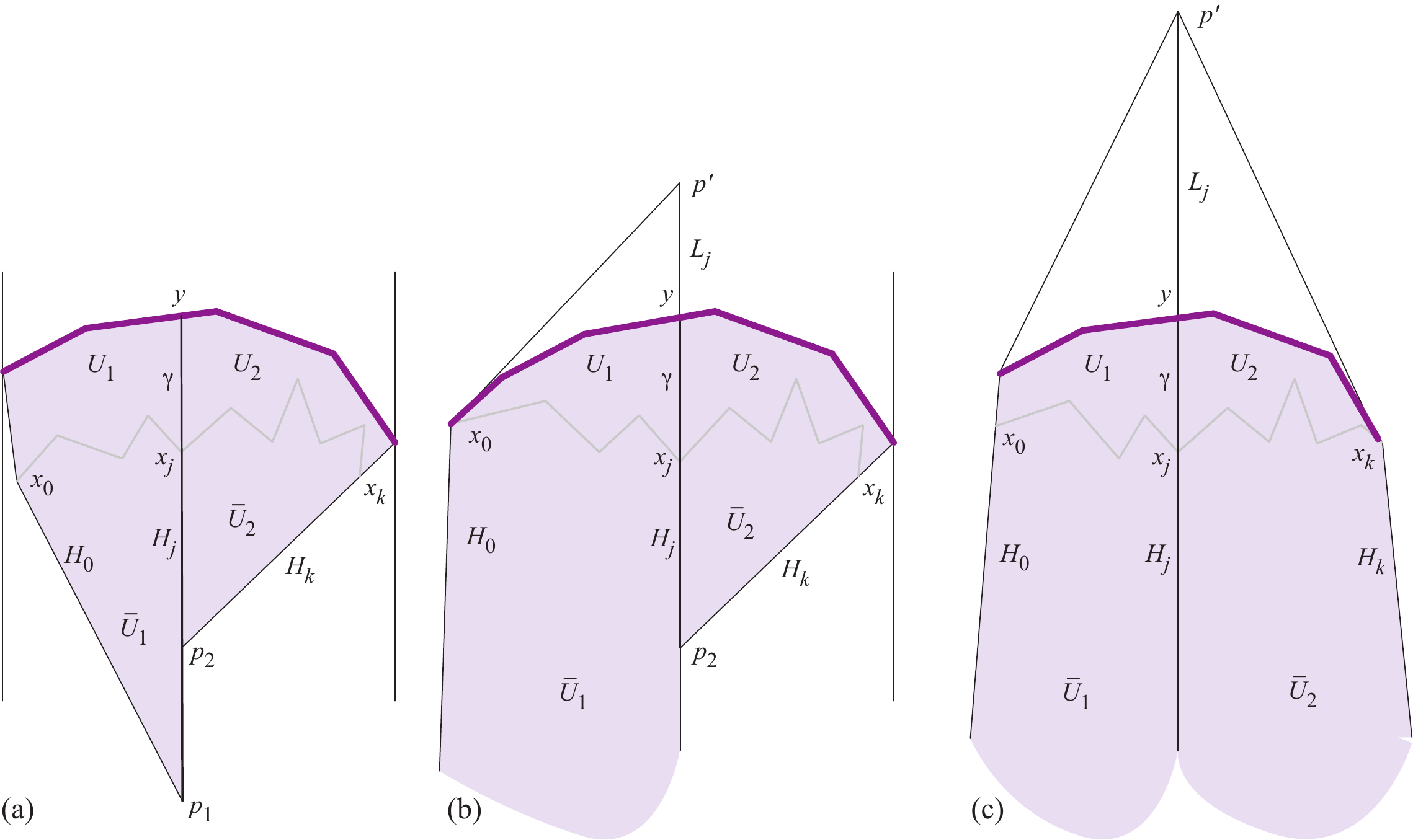}
\caption{Lines determined by the marked portion of $\bU$
are supporting for $U=U_1 \cup U_2$.}
\figlab{Joining}
\end{figure}

For example, $s$ is $v'_0 v'_4$ in Figure~\figref{DigonCube}(b),
and that edge is used to join the halves in Figure~\figref{GeodesicLoopCube}(b).

One issue remains.  It could be that the segment $s$ identified above
is the base of a curvature triangle, rather than a segment of $Q$,
in which case it cannot be used for joining the halves.
Returning to Figure~\figref{Q_convex}, there are two cases:
(a)~$\o < \pi$, and (b)~$\o \ge \pi$.
In case~(a), $s$ is the base of a curvature triangle, with the angle
at either endpoint of the base larger than $\pi/2$.
Note that edges adjacent to $s$ must be edges of $Q$.
Regardless of the angle at which $L_j$ intersects $s$, one of
these two adjacent edges must include supporting portion of $\bU$,
for example, visible from $p'$ in Figure~\figref{Joining}(b,c).
In case~(b), two curvature triangles are inserted,
but as we noted earlier, neither is truly needed, for the boundary
of $\bU$ ($\bP$ in Lemma~\lemref{convex.polygon})
is already convex at the apexes of the two curvature
triangles.  So, simply not inserting them leaves an edge of $Q$
crossed by $L_j$ that can serve as $s$.
So, in all situations, we obtain a supporting segment.

\section{Conclusion}
We have established our main theorem:
\begin{theorem}
Let $Q$ be a quasigeodesic loop on a convex polyhedral surface $\P$.
Cutting shortest paths from every vertex to $Q$, and cutting all but
a supporting segment $s$ of $Q$ as designated above,
unfolds $\P$ to a simple planar polygon.
\thmlab{main}
\end{theorem}

\begin{figure}[htbp]
\centering
\includegraphics[width=0.8\linewidth]{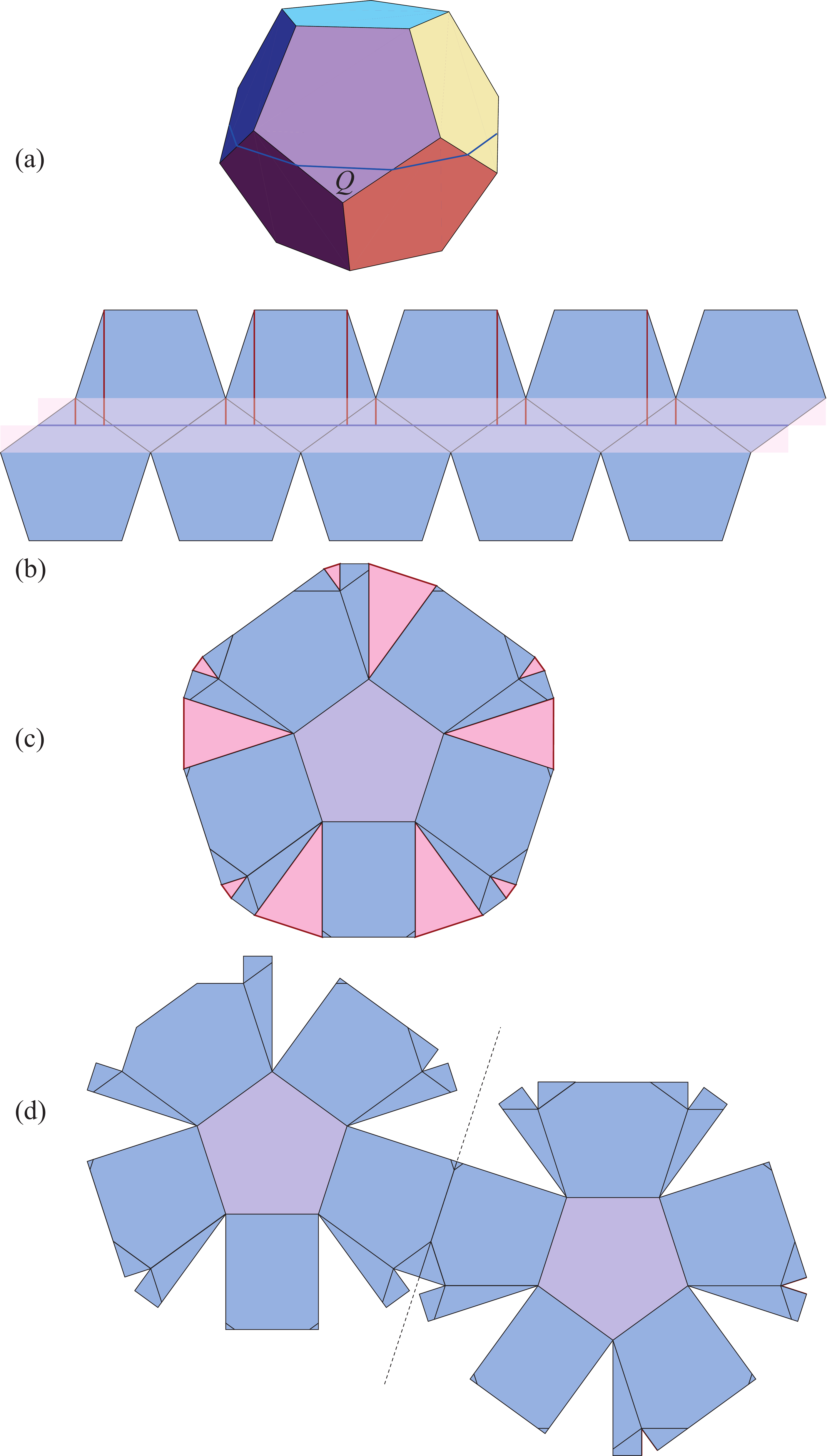}
\caption{(a)~$Q$ here is a geodesic; it includes no vertices, as is
evident in the layout~(b).
The region isometric to a right
circular cylinder is highlighted.
The convex domain $P^\triangle$
from Lemma~\protect\lemref{convex.polygon}
is shown in~(c),
and one possible unfolding in~(d).
}
\figlab{DodecahedronUnfolding}
\end{figure}

Figure~\figref{DodecahedronUnfolding} shows another
example, a closed geodesic on a dodecahedron, this time a pure geodesic.
The unfolding following the above construction is shown
in Figure~\figref{DodecahedronUnfolding}(c,d).
In this case when
$Q$ is a pure, closed geodesic, there is additional structure that
can be used for an alternative unfolding.
For now $Q$
lives on a region isometric to a right
circular cylinder. 
Figure~\figref{DodecahedronUnfolding}(b)
illustrates that
the upper and lower rims of the cylinder are loops
parallel to $Q$ through the vertices of $P$ at minimum distance
to $Q$ (at least one vertex on each side.) 
In the figure, 
these
shortest distances to the upper rim are the short vertical paths from $Q$
to the five pentagon vertices.
Those rim loops are themselves
closed quasigeodesics.
An alternative unfolding keeps the cylinder between the rim loops intact
and attaches the two reduced halves to either side.
See Figure~\figref{DodecaCostin}.
\begin{figure}[htbp]
\centering
\includegraphics[width=0.75\linewidth]{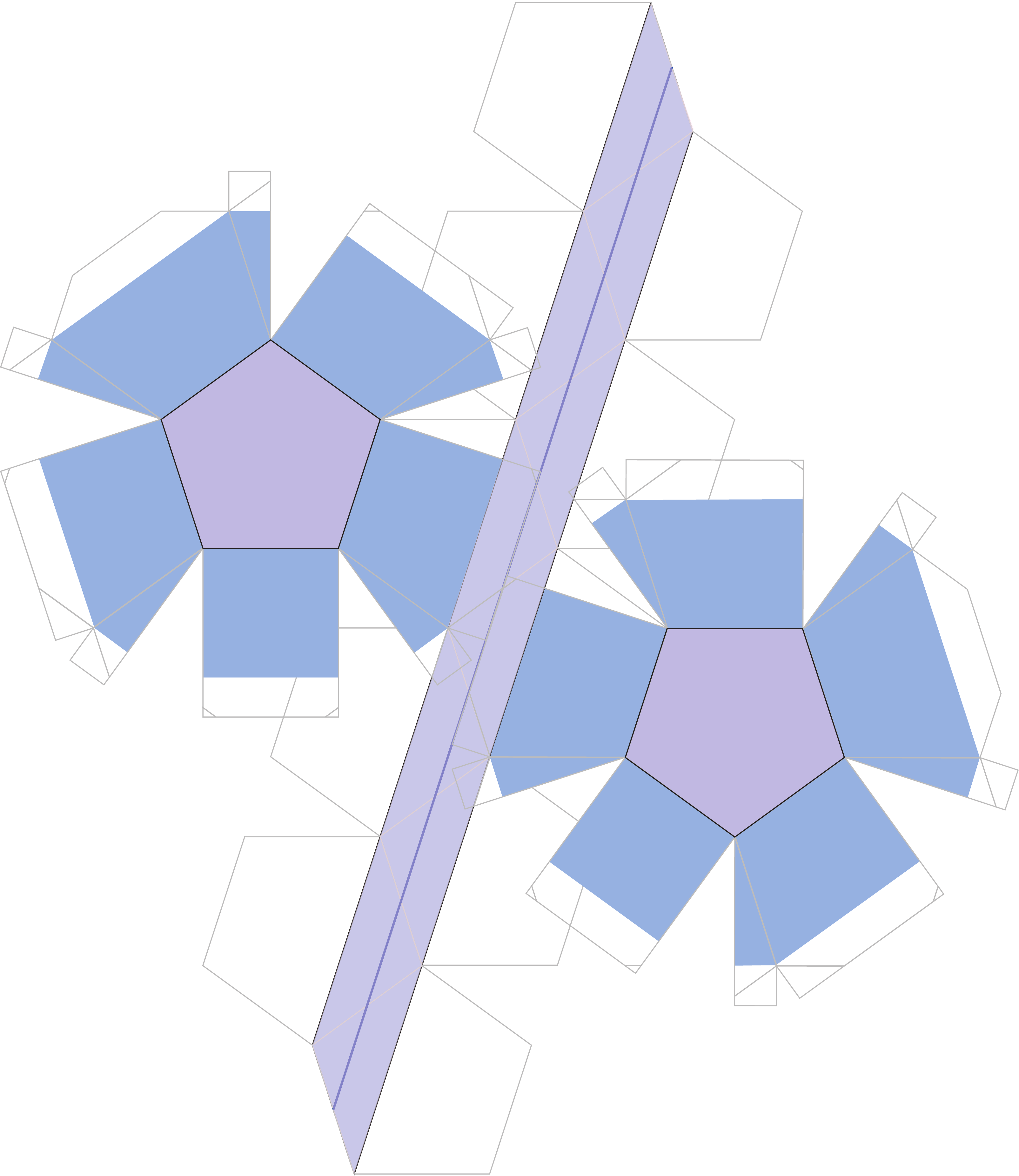}
\caption{Alternative unfolding of the example in
Figure~\protect\figref{DodecahedronUnfolding}.
Various construction lines are shaded lightly.
}
\figlab{DodecaCostin}
\end{figure}

\subsection{Future Work}
We have focused on establishing Theorem~\thmref{main} rather than the algorithmic
aspects.  Here we sketch preliminary thoughts on computational complexity.
Let $n$ be the number of vertices of $\P$, and
let $q=|Q|$ be the number of faces crossed by the
geodesic loop $Q$.
In general $q$ cannot be bound as a function of $n$.
Let $m=n+q$ be the total combinatorial complexity
of the ``input'' to the algorithm.
Constructing $Q$ from a given point and direction will
take $O(q)$ time.
Identifying a supporting segment $s$, and laying out the final
unfolding, is proportional to $m$.
The most interesting algorithmic challenge is to find the shortest
paths from each vertex $v_i$ to $Q$.
The recent linear-time algorithm in~\cite{ss-otasp-08}
leads us to expect the computation can be accomplished efficiently.

We do not believe that quasigeodesic loops constitutes the widest class of
curves for which the star unfolding leads to non-overlap.
Extending Theorem~\thmref{main} to
quasigeodesics with two exceptional points, one with angle larger than $\pi$
to one side, and the other with angle larger than $\pi$ to the other side,
is a natural next step,
not yet completed.

If one fixes a nonvertex point $p \in \P$ and a surface direction $\overrightarrow{u}$ 
at $p$, a quasigeodesic loop 
can be generated to have direction $\overrightarrow{u}$ at $p$.
It might be interesting to study the continuum of star unfoldings 
generated by spinning $\overrightarrow{u}$ around $p$.

\hide{
\section*{Appendix~1: Symbol Glossary}
\begin{tabular}{l l}
$\P$ 
   & convex polyhedron \\
$P_1, P_2$ 
   & the two ``halves'' $\P \setminus Q$ \\
$P$
   & one half, either $P_1$ or $P_2$ \\
$v_i$ 
   & vertex of $\P$ \\
$Q$ 
   & a quasigeodesic loop \\
$x$
   & the exceptional loop point of $Q$ \\
$L(p),R(p)$
   & angle incident to left/right side of $p$ on curve \\
$W$
   & simple, closed curve \\
$\sp(v)$
   & one selected shortest path from $v$ to $Q$ \\
$v'$
   & the projection of $v$ onto $Q$: $\sp(v) = v v'$ \\
$\o=\o(v)$
   & the curvature at $v$, $2\pi$ minus the incident face angles \\
$\triangle$
   & a curvature triangle \\
$\ell$
   & $|\sp(v)|$ \\
$\bP$
   & the boundary of $P$ \\
$P^\triangle$
   & the manifold $P$ after insertion of all curvature triangles $\triangle$ \\
$\o_Q$
   & total curvature inside $Q$ on $P$ \\
$\tau_Q$
   & total turn of $Q = \bP$ \\
$\bP^\triangle$
   & the boundary of  $P^\triangle$ \\
$\b$
   & angle larger than $\pi$ at the loop point $x \in Q$ \\
?
   & \textbf{Large gap here!} \\
$\overrightarrow{u}$ 
   & direction vector through $p \in \P$
\end{tabular}
}

\paragraph{Acknowledgments.}
We thank Boris Aronov for many observations and suggestions 
which improved the paper.

\bibliographystyle{alpha}
\bibliography{/home/orourke/bib/geom/geom}

\begin{thebibliography}{IOV07b}

\bibitem[Ale50]{a-kp-48}
Aleksandr~Danilovich Alexandrov.
\newblock {\em Vupyklue Mnogogranniki}.
\newblock Gosydarstvennoe Izdatelstvo Tehno-Teoreticheskoi Literaturu, 1950.
\newblock In Russian. See~\cite{a-kp-58} for German translation,
  and~\cite{a-cp-05} for English translation.

\bibitem[Ale58]{a-kp-58}
Aleksandr~D. Alexandrov.
\newblock {\em Konvexe Polyeder}.
\newblock Akademie Verlag, Berlin, 1958.
\newblock Math. Lehrbucher und Monographien. Translation of the 1950 Russian
  edition.

\bibitem[Ale05]{a-cp-05}
Aleksandr~D. Alexandrov.
\newblock {\em Convex Polyhedra}.
\newblock Springer-Verlag, Berlin, 2005.
\newblock Monographs in Mathematics. Translation of the 1950 Russian edition by
  N.~S.~Dairbekov, S.~S.~Kutateladze, and A.~B.~Sossinsky.

\bibitem[AO92]{ao-nsu-92}
Boris Aronov and Joseph O'Rourke.
\newblock Nonoverlap of the star unfolding.
\newblock {\em Discrete Comput. Geom.}, 8:219--250, 1992.

\bibitem[AZ67]{az-igs-67}
Aleksandr~D. Alexandrov and Victor~A. Zalgaller.
\newblock {\em Intrinsic Geometry of Surfaces}.
\newblock American Mathematical Society, Providence, RI, 1967.

\bibitem[DO07]{do-gfalop-07}
Erik~D. Demaine and Joseph O'Rourke.
\newblock {\em Geometric Folding Algorithms: Linkages, Origami, Polyhedra}.
\newblock Cambridge University Press, July 2007.
\newblock \url{http://www.gfalop.org}.

\bibitem[IIV07]{iiv-qfpcs-07}
Kouki Ieiri, {Jin-ichi} Itoh, and Costin V\^{i}lcu.
\newblock Quasigeodesics and farthest points on convex surfaces.
\newblock Submitted, 2007.

\bibitem[IOV07a]{iov-ucpq-07}
{Jin-ichi} Itoh, Joseph O'Rourke, and Costin V\^{i}lcu.
\newblock Unfolding convex polyhedra via quasigeodesics.
\newblock Technical Report 085, Smith College, July 2007.
\newblock arXiv:0707.4258v2 [cs.CG].

\bibitem[IOV07b]{iov-ucpq-07a}
{Jin-ichi} Itoh, Joseph O'Rourke, and Costin V\^{i}lcu.
\newblock Unfolding convex polyhedra via quasigeodesics: {A}bstract.
\newblock In {\em Proc. 17th Annu. Fall Workshop Comput. Comb. Geom.}, November
  2007.

\bibitem[IOV09]{iov-sucpr-09}
{Jin-ichi} Itoh, Joseph O'Rourke, and Costin V\^{i}lcu.
\newblock Source unfoldings of convex polyhedra with respect to certain closed
  polygonal curves.
\newblock In {\em Proc. 25th European Workshop Comput. Geom.}, pages 61--64.
  EuroCG, March 2009.

\bibitem[IV08a]{iv-cfpcs-08}
{Jin-ichi} Itoh and Costin V\^{i}lcu.
\newblock Criteria for farthest points on convex surfaces.
\newblock \emph{Mathematische Nachrichten}, to appear, 2008.

\bibitem[IV08b]{iv-gcit-08}
{Jin-ichi} Itoh and Costin V\^{i}lcu.
\newblock Geodesic characterizations of isosceles tetrahedra.
\newblock Preprint, 2008.

\bibitem[Kob67]{k-ccl-67}
Shoschichi Kobayashi.
\newblock On conjugate and cut loci.
\newblock In S.~S. Chern, editor, {\em Studies in Global Geometry and
  Analysis}, pages 96--122. Mathematical Association of America, 1967.

\bibitem[MP08]{mp-mccpc-05}
Ezra Miller and Igor Pak.
\newblock Metric combinatorics of convex polyhedra: {C}ut loci and
  nonoverlapping unfoldings.
\newblock {\em Discrete Comput. Geom.}, 39:339--388, 2008.

\bibitem[OV09]{ov-npsuc-09}
Joseph O'Rourke and Costin V\^{i}lcu.
\newblock A new proof for star unfoldings of convex polyhedra.
\newblock Manuscript in preparation, 2009.

\bibitem[Pog49]{p-qglcs-49}
Aleksei~V. Pogorelov.
\newblock Quasi-geodesic lines on a convex surface.
\newblock {\em Mat. Sb.}, 25(62):275--306, 1949.
\newblock English transl., {\em Amer. Math. Soc. Transl.} 74, 1952.

\bibitem[Pog73]{p-egcs-73}
Aleksei~V. Pogorelov.
\newblock {\em Extrinsic Geometry of Convex Surfaces}, volume~35 of {\em
  Translations of Mathematical Monographs}.
\newblock American Mathematical Society, Providence, RI, 1973.

\bibitem[Poi05]{p-lgsc-1905}
Henri Poincar\'e.
\newblock Sur les lignes g\'eod\'esiques des surfaces convexes.
\newblock {\em Trans. Amer. Math. Soc.}, 6:237--274, 1905.

\bibitem[Sak96]{s-rg-96}
Takashi Sakai.
\newblock {\em Riemannian Geometry}.
\newblock Translation of Mathematical Monographs 149. Amer. Math. Soc., 1996.

\bibitem[SS86]{ss-spps-86}
Micha Sharir and Amir Schorr.
\newblock On shortest paths in polyhedral spaces.
\newblock {\em SIAM J. Comput.}, 15:193--215, 1986.

\bibitem[SS08]{ss-otasp-08}
Yevgeny Schreiber and Micha Sharir.
\newblock An optimal-time algorithm for shortest paths on a convex polytope in
  three dimensions.
\newblock {\em Discrete \& Comput. Geom.}, 39:500--579, 2008.

\bibitem[ST96]{st-cldsa-96}
Katsuhiro Shiohama and Minoru Tanaka.
\newblock Cut loci and distance spheres on alexandrov surfaces.
\newblock {\em S{\'e}minaires \& Congr{\`e}s}, 1:531--559, 1996.
\newblock Actes de la Table Ronde de G{\'e}om{\'e}trie Diff{\'e}rentielle
  (Luminy, 1992).

\end{thebibliography}
\end{document}